%% file: main.tex
\documentclass[ 
reprint,
 amsmath,amssymb,
 aps,
 pra,
floatfix,
]{revtex4-2}

\usepackage{orcidlink}
\usepackage{subfiles}
\usepackage{tikz}
\usetikzlibrary{decorations.pathmorphing,patterns}
\usepackage{pgfplots}
\pgfplotsset{compat=1.17} 
\usepgfplotslibrary{dateplot}

\usepackage{subfigure}
\usepackage{graphicx}
\usepackage{dcolumn}
\usepackage{amsmath,amsfonts}
\usepackage{bm}

\begin{document}

\title{Dynamical Networking using Gaussian fields}

\author{Nadine du Toit \orcidlink{0000-0003-0262-7010}}
\email{Corresponding author: 24461989@sun.ac.za}
\affiliation{
Department of Physics, Stellenbosch University, Stellenbosch 7602, South  Africa\\
}
 
\author{Kristian K. M\"uller-Nedebock \orcidlink{0000-0002-1772-1504}}
 \email{kkmn@sun.ac.za}
\affiliation{
Department of Physics, Stellenbosch University, Stellenbosch 7602, South  Africa
}
\affiliation{National Institute for Theoretical and Computational Sciences, Stellenbosch, 7602, South Africa}

\date{\today}

\begin{abstract}

A novel field theoretical approach towards modelling dynamic networking in complex systems is presented. An equilibrium networking formalism which utilises Gaussian fields is adapted to model the dynamics of particles that can bind and unbind from one another. Here, \textit{networking} refers to the introduction of instantaneous co-localisation constraints and does not necessitate the formation of a well-defined transient or persistent network. By combining this formalism with Martin-Siggia-Rose generating functionals, a weighted generating functional for the networked  system is obtained. The networking formalism introduces spatial and temporal constraints into the Langevin dynamics, via statistical weights, thereby accounting for all possible configurations in which particles can be networked to one another. A simple example of Brownian particles which can bind and unbind from  one another demonstrates the tool and that this leads to results for physical quantities in a collective description. Applying the networking formalism to model the dynamics of cross-linking polymers in a mixture, we can calculate the average number of networking instances. As expected, the dynamic structure factors for each type of polymer show that the system collapses once networking is introduced, but that the addition of a repulsive time-dependent potential above a minimum strength prevents this. The examples presented in this paper indicate that this novel approach towards modelling dynamic networking could be applied to a range of synthetic and biological systems to obtain theoretical predictions for experimentally verifiable quantities. 
\end{abstract}

\maketitle

\section{Introduction}
\label{sec:intro}

In the last decades, both biological and synthetic cross-linked polymer networks have attracted significant interest across various disciplines, from the fundamental study of their mechanisms \cite{luFormationMechanismHydrogels2018,katashimaRheologicalStudiesPolymer2021}, to their  applications in biotechnology \cite{zhangRecentAdvancesEnvironmentally2024} and materials design \cite{webberDynamicReconfigurableMaterials2022}. The cytoskeleton is a primary example of such a polymer network. Various other biological processes, such as transport of vesicles by molecular motors, also involve proteins and molecules binding and unbinding to one another as they move throughout the intracellular environment. In the context of the cytoskeleton and its dynamics molecular motors can act as  active cross-linkers, not only binding cytoskeletal filaments to one another, but also introducing forces and initiating motion on length scales of the order of the whole cell \cite{gongCrosslinkedBiopolymerNetworks2019}. Cross-linked biopolymers present promising avenues of exploration for the development of biotechnology and biomedical applications \cite{reddyCrosslinkingBiopolymersBiomedical2015,khanEngineering3DPolymer2022}. 

Quantitative knowledge of these dynamical systems has increased significantly in the last few decades due to sophisticated imaging techniques \cite{finkenstaedt-quinnSuperresolutionImagingMonitoring2016,chaubetDynamicActinCrosslinking2020} as well as simulations \cite{garduno-juarezMolecularDynamicSimulations2024}. Various theoretical modelling approaches  have lead to significant insights on the microscopic dynamics of single filaments \cite{mackintoshPolymerbasedModelsCytoskeletal2001,Doi&EdwardsBook}. Coarse-grained and continuum models have lead to further insights on network properties and solutions of polymers \cite{broederszModelingSemiflexiblePolymer2014a,Furthauer2022}. Despite these significant advancements, modelling the complex dynamics involving binding and unbinding of particles exhibited in biological systems such as the cytoskeleton, remains challenging. The complex behaviour of these biological systems suggest the need for a deeper understanding of their dynamics and the the effects of introducing microscopic spatio-temporal constraints --via the binding and unbinding of particles, on the collective dynamics of a system. This paper presents an abstract modelling approach towards generic systems of particles binding and unbinding to one another.
The formalism allows for various possible implementations which can be adjusted to model aspects of a range of synthetic and biological systems. 

A networking formalism presented by Edwards \cite{Boue, Edwards1988},  models cross-linked polymers in equilibrium on the mesoscopic scale. This paper presents a novel modelling approach by introducing adaptations into Edwards' networking formalism in order to model dynamical networking in systems where particles bind and unbind to another over time.  It should be noted that \textit{networking} and \textit{networked}, in this context, refer to the introduction of spatio-temporal constraints into the system to model the binding and unbinding of particles. While this may lead to network-like behaviour for some instances, introducing the networking formalism into a system does not necessarily result in a well-defined transient or persistent network. Since the networking formalism introduces constraints, it must be interpreted within the context of a system's dynamics. In this paper, it is combined with the Martin-Siggia-Rose (MSR) generating functional method \cite{Martin:1973zz,Jensen1981a}, which has been successfully applied to polymeric systems \cite{rostiashviliLangevinDynamicsGlass1998,fredricksonCollectiveDynamicsPolymer1990}, thereby enabling the incorporation of spatial and temporal constraints into the Langevin dynamics of the systems under consideration. The formalism incorporates these constraints by accounting for all possible configurations in which these constraints can be implemented. This approach, along with the proposed approximation schemes, yields results for the networked system in a collective description wherein the networking acts as time-dependent effective potentials between the particles. Experimentally accessible quantities, such as dynamic structure factors for the networked system, are derived. In the proposed approach the use of this formalism is limited to dynamical systems with small density fluctuations, but this limitation is not inherent to the  networking formalism itself, rather a consequence of the approximation schemes used in  modelling of the collective dynamics of the systems considered here.

The paper begins by introducing the reader to the networking formalism as utilised in equilibrium \cite{Edwards1988} in Sect.~\ref{sec:networking}. Adaptations are introduced into the theory in Sect.~\ref{sec:DynamicalNetworking} to allow for dynamical networking. The proposed approximations schemes are implemented and discussed in the context of an illustrative example where Brownian particles are dynamically networked to one another. The subsequent discussion introduces further refinements into the networking formalism which is finally applied to model the scenario of cross-linking in a polymer mixture.

\section{The networking theory}
\label{sec:networking}

\begin{figure}
\begin{center}
\begin{tikzpicture}[scale = 0.8]
	
		\node [] (3) at (-1, 0.5) {};
		\node [] (4) at (-1.5, 1) {};
		\node [] (5) at (-0.5, 1) {};
		\node [] (21) at (-1, 0) {};
		\node [] (45) at (1, 0.5) {};
		\node [] (48) at (1, 0) {};
		\node [] (66) at (-2, 0) {};
		\node [] (67) at (8, 0) {};
		\node [] (68) at (1, 0.5) {};
		\node [] (69) at (0.5, 1) {};
		\node [] (70) at (1.5, 1) {};
		\node [] (71) at (3, 0.5) {};
		\node [] (72) at (3, 0) {};
		\node [] (73) at (3, 0.5) {};
		\node [] (74) at (2.5, 1) {};
		\node [] (75) at (3.5, 1) {};
		\node [] (76) at (5, 0.5) {};
		\node [] (77) at (5, 0) {};
		\node [] (78) at (5, 0.5) {};
		\node [] (79) at (4.5, 1) {};
		\node [] (80) at (5.5, 1) {};
		\node [] (81) at (7, 0.5) {};
		\node [] (82) at (7, 0) {};
		\node [] (83) at (7, 0.5) {};
		\node [] (84) at (6.5, 1) {};
		\node [] (85) at (7.5, 1) {};

		\node [red] at (-1,4) {\large $\mathbf{r}_1'$};
		\draw  node[fill= red,circle,scale=2.5,above] (86) at (-1, 2.5) {};
		
		\node [red] at (1,4) {\large $\mathbf{r}_2'$};
		\draw  node[fill= red,circle,scale=2.5,above]  at (1, 2.5) {};
		
		\node [red] at (3,4) {\large $\mathbf{r}_3'$};
		\draw  node[fill= red,circle,scale=2.5,above]  at (3, 2.5) {};
		
		\node [red] at (5,4) {\large $\mathbf{r}_4'$};
		\draw  node[fill= red,circle,scale=2.5,above]  at (5, 2.5) {};
		
		\node [red] at (7,4) {\large $\mathbf{r}_5'$};
		\draw  node[fill= red,circle,scale=2.5,above]  at (7, 2.5) {};

		\draw [blue, ultra thick] (3.center) to (21.center) node[below]{\large $\mathbf{r}_1$};
		\draw [blue, ultra thick, bend right=90, looseness=1.75] (4.center) to (5.center);
		\draw [blue, ultra thick] (45.center) to (48.center) node[below]{\large $\mathbf{r}_2$};

		\draw [blue, ultra thick, bend right=90, looseness=1.75] (69.center) to (70.center);
		\draw [blue, ultra thick] (71.center) to (72.center) node[below]{\large $\mathbf{r}_3$};
		\draw [blue, ultra thick, bend right=90, looseness=1.75] (74.center) to (75.center);
		\draw [blue, ultra thick] (76.center) to (77.center) node[below]{\large $\mathbf{r}_4$};
		\draw [blue, ultra thick, bend right=90, looseness=1.75] (79.center) to (80.center);
		\draw [blue, ultra thick] (81.center) to (82.center) node[below]{\large $\mathbf{r}_5$};
		\draw [blue, ultra thick, bend right=90, looseness=1.75] (84.center) to (85.center);
	
\end{tikzpicture}
\end{center}
    \caption{Schematic diagram of \textit{beads} and \textit{cups} representing \textit{attachment points} to be paired with the networking theory.}
    \label{fig:EquilibriumNetworking}
\end{figure}
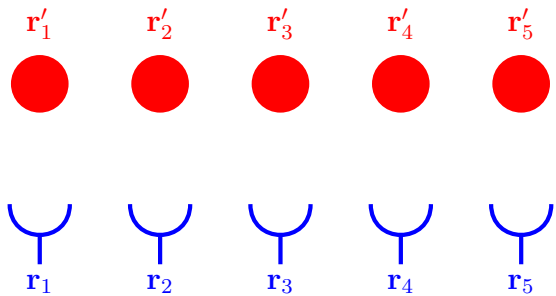

Fig.~\ref{fig:EquilibriumNetworking} depicts two sets of variables, namely  $\{r_1,r_2,r_3,r_4,r_5\}$ and $\{r_1',r_2',r_3',r_4',r_5'\}$. These variables will be utilised to illustrate how the networking theory enumerates pairs of variables, whilst equating all possible pairs and pairings. For this example, the sets of variables have been chosen as spatial coordinates, but could certainly have been of another variety. The set of $r_i'$ have been chosen to represent the coordinates of five \textit{beads}, whilst the $r_i$ represent five points to which these beads may attach or \textit{network}. The theory provides an expression which spatially constrains each bead to an \textit{attachment point}, simultaneously accounting for all possible ways in which the five beads may attach to the five points. In the discussion to follow, the reader will be introduced to some generalised Gaussian path integrals and guided through the process in which such a networking expression may be constructed from these integrals.

Following the notation from Edwards \cite{Edwards1988}, consider the Gaussian integrals

\begin{equation}
    \frac{\int \Pi \mathrm{d} \Phi \, \Phi_n \Phi_m \, \mathrm{e}^{-\frac{1}{2}\sum_i \sum_j \Phi_i (a_{ij})^{-1} \Phi_j} }{\int \Pi \mathrm{d} \Phi \,  \mathrm{e}^{-\frac{1}{2}\sum_i \sum_j \Phi_i (a_{ij})^{-1} \Phi_j} } =
  a_{n m} \,,
  \end{equation}
  \begin{eqnarray}
      \frac{\int \Pi \mathrm{d} \Phi \, \Phi_a \Phi_b\Phi_c \Phi_d \, \mathrm{e}^{-\frac{1}{2}\sum_i \sum_j \Phi_i (a_{ij})^{-1} \Phi_j} }{\int \Pi \mathrm{d} \Phi \,  \mathrm{e}^{-\frac{1}{2}\sum_i \sum_j \Phi_i (a_{ij})^{-1} \Phi_j} } \nonumber\\
 = a_{a b}a_{c d}+ a_{a c}a_{b d}+ a_{a d}a_{b c} \,.
\end{eqnarray}
where the  $a_{ij}$ are elements of a positive-definite matrix. As explained by Edwards\cite{Edwards1988},  the analogous continuum expressions follow simply from the above if there are many instances of $\Phi_n$ and $\Phi_m$ that are separated by uniform intervals. The product of integrals may then be replaced by a functional integral and the Kronecker delta by a Dirac delta. In addition, complex notation may be utilised such that the fields $\Phi$ and $\Phi^*$ are given in terms of two independent real fields $\Phi_1$ and $\Phi_2$ as $\Phi = \Phi_1 + \mathrm{i}\Phi_2$ and  $\Phi^* = \Phi_1 - \mathrm{i}\Phi_2$. Introducing the shorthand notation $ \int_r= \int_{-\infty }^{\infty} \mathrm{d} r$ for the integrals in the exponent and utilising square braces to denote path integrals, the Gaussian integrals for these fields become:
\begin{subequations}
\begin{eqnarray}
\int [ \mathrm{d} \Phi] [\mathrm{d} \Phi^*]
\,\Phi(r) \Phi(r') \, \mathrm{e}^{-\int_r \Phi(r) \,\Phi^*(r)} =0\, , \\
\int [ \mathrm{d} \Phi] [\mathrm{d} \Phi^*]\, \Phi^*(r) \Phi^*(r') \, \mathrm{e}^{-\int_r \Phi(r)\, \Phi^*(r)} =0
\end{eqnarray}
for pairs of $\Phi$ and for pairs of $\Phi^*$ and 
\begin{equation}
  \mathcal{N}\int [ \mathrm{d} \Phi] [\mathrm{d} \Phi^*]\, \Phi(r) \,\Phi^*(r') \, \mathrm{e}^{-\int_r \Phi(r)\, \Phi^*(r)} =  \delta (r-r')\, ,
\end{equation}
\label{eq:GaussianIntegrals}
\end{subequations}
for pairs of $\Phi$ and $\Phi^*$. Here $\mathcal{N}$ is a normalisation given by
\begin{equation}
    \mathcal{N}^{-1} = \int [ \mathrm{d} \Phi] [\mathrm{d} \Phi^*]  \, \mathrm{e}^{-\int_r \Phi(r)\, \Phi^*(r)}\, .
\end{equation}
With this formulation, all integrals containing odd multiples of the fields $\Phi $ and $\Phi^*$ vanish, \textit{e.g.},
\begin{equation}
    \int [ \mathrm{d} \Phi] [\mathrm{d} \Phi^*]
\,\Phi(r) \Phi^*(r') \Phi^*(r'') \, \mathrm{e}^{-\int_r \Phi(r) \,\Phi^*(r)} =0\, .
\end{equation}
Integrals with two or more pairs of $\Phi$ and $\Phi^*$ will evaluate to yield multiple terms \textit{e.g.}
\begin{align}
 \mathcal{N} &\int [ \mathrm{d} \Phi] [\mathrm{d} \Phi^*]\, \Phi(r_1) \, \Phi(r_2) \, \Phi^*(r'_1) \,\Phi^*(r'_2) \,  \mathrm{e}^{-\int_r \Phi(r)\, \Phi^*(r)}\,  \nonumber \\
  &=\delta (r_1-r'_1) \, \delta (r_2-r'_2)+ \delta (r_2-r'_1) \, \delta (r_1-r'_2)\,
  \end{align}
such that all possible configurations of paired-off variables are accounted for.
Still borrowing the notation from \cite{Edwards1988}, the fields $\Phi$ and $\Phi^*$ evaluated at points $r_1,..., r_n$ and  $r'_1,..., r'_m$, may be used to construct the following integral
\begin{eqnarray}
  I =\mathcal{N}\int [ \mathrm{d} \Phi] [\mathrm{d} \Phi^*]\, \big\{ \Phi(r_1) \, \Phi(r_2) \,...\,\Phi(r_n)\nonumber \\
  \times \Phi^*(r'_1) \,\Phi^*(r'_2) \,...\,\Phi^*(r'_m)  \mathrm{e}^{-\int_r \Phi(r)\, \Phi^*(r)}\big\}\, .
  \label{eq:networkingExample} 
\end{eqnarray}
If $n=m$ this yields 
\begin{equation}
I =\sum_\mathrm{pairings}   \left(\prod_\mathrm{pairs} \delta(r_i-r'_j) \right)\, ,
\label{eq:pairings}
\end{equation}
otherwise the integral equals $0$. Applying this to the five beads and attachment points in Fig.~\ref{fig:EquilibriumNetworking}, we have that $n=m=5$ such that the path integrals in ~eq.~\eqref{eq:networkingExample} evaluate to ~eq.~\eqref{eq:pairings}. In this case, each term in the sum would be a product of five Dirac delta functions that pair each bead's position with the position of an attachment point, whilst accounting for all possible arrangements of these pairings with the sum itself. Furthermore, eq.~\eqref{eq:pairings} shows that the integral not only accounts for all possible pairings but also ensures that the variables in a pair have the same value. The key thing to note in order to clarify this statement, is that ~eq.~\eqref{eq:networkingExample} is designed to be utilised within another formalism. One would, for example, introduce ~eq.~\eqref{eq:networkingExample} into an already existing set of expressions describing the equilibrium behaviour of the positions of the beads and/or attachment points. This would impose an additional set of constraints on the system, requiring that the positions of the beads coincide with those of the attachment points. Since ~eq.~\eqref{eq:networkingExample} networks pairs of variables in this manner, expressions of this form will be referred to as \textit{networking functionals}. It is worth noting that networking, in this context, again refers to the spatial and temporal constraints that are introduced and does not necessarily result in the formation of spatially spanning networks.

By judiciously combining expressions analogous to eqs.~\eqref{eq:GaussianIntegrals}, a networking functional may be constructed which pairs off spatial, as well as temporal variables. This will be demonstrated (see ~Sect.~\ref{sec:DynamicalNetworking}) by introducing an additional dependence on a temporal variable into the field $\Phi(r,t)$ and its complex conjugate $\Phi^*(r,t)$ in such a manner that the set of beads may periodically attach and detach from the set of attachment points.

\section{Dynamical networking}
\label{sec:DynamicalNetworking}

Consider $N$ beads with positions $\mathcal{B}=\{\mathbf{r'_1}(t),\mathbf{r'_2}(t), ... ,\mathbf{r'_N}(t) \}$ and $M$ \textit{attachment points} with positions $\mathcal{A}=\{\mathbf{r_1}(t),\mathbf{r_2}(t), ... ,\mathbf{r_M}(t) \}$, similar to the depiction in Fig. \ref{fig:EquilibriumNetworking}, with the addition of time-dependence to account for any motion of the beads and attachment points. To consecutively network all of the beads to the attachment points at incremental time steps, $t_1,t_2, t_3, ...$ evenly separated by a time interval $\tau$, we may set up the following networking functional:
\begin{widetext}
    \begin{equation}
    \int [ \mathrm{d}\Phi] [\mathrm{d} \Phi^*]\,\, \prod_j \left( \prod_{m=1}^M (1+\Phi(\mathbf{r_m},t_j))
    \, \prod_{n=1}^N \Phi^*(\mathbf{r'_n},t_j) \, \mathrm{e}^{-  \int_\mathbf{y} \, \Phi(\mathbf{y},t_j)\,\Phi^*(\mathbf{y},t_j)} \right)\,
        \label{eq:networkingConstraintRaw}    
    \end{equation}
\end{widetext}
It should be noted that this expression only holds for scenarios where there are at least as many attachment points as there are beads, \textit{i.e.} $N \leq M$. This restriction arises due to the $1+ \Phi$ in the product over $m$, which allows for the attachment points to remain non-networked whilst requiring that all beads  must be networked to an attachment site. This is, however, an arbitrary choice which has been made and can  be adjusted to model alternative scenarios by reconfiguring the products in \eqref{eq:networkingConstraintRaw}. Collective variables
\begin{subequations}
\begin{equation}
    \rho_\mathrm{B}(\mathbf{r},t) = \sum_{n=1}^N \delta(\mathbf{r}-\mathbf{r'_n}(t)) \, 
    \label{eq:rhoB}
\end{equation}
for the positions of the beads and
\begin{equation}
    \rho_\mathrm{A}(\mathbf{r},t) = \sum_{m=1}^M \delta(\mathbf{r}-\mathbf{r_m}(t)) \, 
    \label{eq:rhoA}
\end{equation}
\end{subequations}
for the positions of the attachment points, may be introduced such that the networking functional may be rewritten in a continuous representation. For example $\rho_\mathrm{A}(\mathbf{r},t)$ may be introduced as follows:
\begin{align}
\Pi_j\Pi_n(1+\Phi(r_n,t_j)) &{}= \mathrm{e} ^{\Sigma_j\Sigma_n\mathrm{ln}(1+\Phi(r_n,t_j))}\\
&{}= \mathrm{e} ^{\Sigma_j\int_{r}\rho_\mathrm{A}(r,t_j)\mathrm{ln}(1+\Phi(r,t_j))}.
\label{eq:exponentiation}
\end{align}
Taking the continuum limit of the time discretisation, the sum over $j$  becomes an integral such that
\begin{equation}
\mathrm{e} ^{\Sigma_j\int_{r}\rho(r,t_j)\mathrm{ln}(1+\Phi(r,t_j))}\rightarrow\mathrm{e} ^{\frac{1}{\tau}\int_{r,t}\rho(r,t)\mathrm{ln}(1+\Phi(r,t))}\,.
\label{eq:exponentiationContinuum}
\end{equation}
Here $\tau$ arises to ensure that the argument of the exponent is dimensionless and gives the constant time interval separating the times $t_j$ in the discretisation of the networking functional.  Applying this to all terms of the networking functional one obtains
\begin{equation}
 \mathbb{Q}[\rho_\mathrm{B}, \rho_\mathrm{A}]=\mathcal{N}_\Phi \int [ \mathrm{d}\Phi] [\mathrm{d} \Phi^*]\,\, \mathrm{e}^ {\mathcal{F}[\Phi, \Phi^*] }\,,
\label{eq:networkingConstraint}    
\end{equation}
where
\begin{eqnarray}
    \mathcal{F}[\Phi, \Phi^*]=  \frac{1}{\tau}\int_{\mathbf{r},t} \rho_\mathrm{A}(\mathbf{r},t)\mathrm{ln}(1+\Phi(\mathbf{r},t))\nonumber\\
    +\frac{1}{\tau}\int_{\mathbf{r},t} \rho_\mathrm{B}(\mathbf{r},t)\mathrm{ln}
\, \Phi^*(\mathbf{r},t) \, \nonumber\\
-  \alpha \int_{\mathbf{y},t} \, \Phi(\mathbf{y},t)\,\Phi^*(\mathbf{y},t)
\label{eq:FbeforeSP}
\end{eqnarray}
and 
 \begin{equation}
     \mathcal{N}_\Phi^{-1}  = \int [ \mathrm{d}\Phi] [\mathrm{d} \Phi^*]\,\, \mathrm{e}^{- \alpha \int_{\mathbf{y},t} \, \Phi(\mathbf{y},t)\,\Phi^*(\mathbf{y},t)}\,
 \end{equation}
 gives the normalisation. The additional constant $\alpha$  is included to ensure that the argument of the exponent remains dimensionless and is discussed in more detail in Sec.~\ref{Sec:BrownianExample}, where it is also shown to fall away in subsequent approximations. The nonlinear nature of the functional integrals over $\Phi$ and $\Phi^\star$ in eq.~\eqref{eq:networkingConstraint}  suggests the use of a saddle-point approximation.  \\

\subsection{The Saddle Point Approximation}
\label{subsec:saddlepoint}
 Determining the extrema of $\mathcal{F}$ in the exponent through
\begin{subequations}
    \begin{eqnarray}
0=\left.\frac{\partial \mathcal{F}}{\partial \Phi(r,t)} \right|_{\bar{\Phi}^* ,\bar{\Phi}} \label{eq:SP1} \\
0=\left.\frac{\partial \mathcal{F}}{\partial \Phi^*(r,t)} \right|_{\bar{\Phi}^* ,\bar{\Phi}}\label{eq:SP2}
\end{eqnarray}
\end{subequations}
means that 

up to lowest order, this approximation can be written as

\begin{equation}
\mathbb{Q}[\rho_\mathrm{B}, \rho_\mathrm{A}]=\mathcal{N}_\Phi  \mathrm{e}^{\mathcal{F}[\bar{\Phi}, \bar{\Phi}^*]} \,,
\label{eq:SPApprox}    
\end{equation}
Referring back to ~eq.~\eqref{eq:FbeforeSP}, each term in the exponent of ~eq.~\eqref{eq:SPApprox} is of the order of either the number of beads or attachment points. If  $\mathcal{F}$ is of the order of the number of particles or higher order, as is the case here, the saddle point approximation yields physical results in the thermodynamic limit.

Solving the simultaneous eqs.(\ref{eq:SP1}) and (\ref{eq:SP2})  yields 
\begin{eqnarray}
    \bar{\Phi}(\mathbf{r},t) = \frac{\rho_\mathrm{B}(\mathbf{r},t)}{\rho_\mathrm{A}(\mathbf{r},t)-\rho_\mathrm{B}(\mathbf{r},t)} \label{eq:SPsol1}\, ,\\
    \bar{\Phi}^*(\mathbf{r},t) = \frac{\rho_\mathrm{A}(\mathbf{r},t)-\rho_\mathrm{B}(\mathbf{r},t)}{\tau \alpha}\label{eq:SPsol2}\, .
\end{eqnarray}
In order for $\bar{\Phi}^*$ to be dimensionless $\tau\alpha= \tfrac{1}{\ell }$ where $\ell$   is a length scale associated with networking. In practice this length scale becomes irrelevant in further approximation schemes, but should in principle be representative of the distance particles of $\mathrm{A}$ and $\mathrm{B}$  are expected to travel in a time $\tau$.  Substituting eqs.~\eqref{eq:SPsol1}-\eqref{eq:SPsol2} , the saddle point approximation for the networking functional is  given by
\begin{widetext}
    \begin{equation}
    \mathbb{Q}[\rho_\mathrm{B},\rho_\mathrm{A}]=\mathcal{N}_\Phi  \mathrm{e}^{\frac{1}{\tau}\int_{\mathbf{r},t} \rho_\mathrm{A}(\mathbf{r},t)\mathrm{ln}\left(1+\frac{\rho_\mathrm{B}(\mathbf{r},t)}{\rho_\mathrm{A}(\mathbf{r},t)-\rho_\mathrm{B}(\mathbf{r},t)} \right)+\frac{1}{\tau}\int_{\mathbf{r},t} \rho_\mathrm{B}(\mathbf{r},t)\mathrm{ln}(\frac{\rho_\mathrm{A}(\mathbf{r},t)-\rho_\mathrm{B}(\mathbf{r},t)}{\tau \alpha} )\, -  \frac{1}{\tau} \int_{\mathbf{y},t} \, \rho_\mathrm{B}(\mathbf{y},t)}\, .
        \label{eq:SPnetworking}    
    \end{equation}
\end{widetext}
As will be discussed in Sect.~\ref{Sec:BrownianExample}, with the aid of an illustrative example, the Saddle Point approximation in ~eq.~\eqref{eq:SPnetworking} leads to a clear physical interpretation in both microscopic as well as collective descriptions of the system.
\subsection{Example: Networked Brownian particles}
\label{Sec:BrownianExample}
To illustrate, consider the scenario where each of the sets of collective variables, $ \rho_\mathrm{A} (\mathbf{r},t)$ and $\rho_\mathrm{B}( \mathbf{r},t)$, describe the dynamics of a set of Brownian particles. The overdamped Langevin equations for each of the beads with position  $\mathbf{r}_\mathrm{B}$ and attachment points with position $\mathbf{r}_j$ can be written as
\begin{subequations}
\label{eq:LangevinBM}
\begin{eqnarray}
\mathcal{L}_\mathrm{A} = - \gamma \dot{\mathbf{r}}_\mathrm{A} (t) + \mathbf{f}_\mathrm{A} (t) =0 \, , \\
\mathcal{L}_\mathrm{B} = - \gamma \dot{\mathbf{r}}_\mathrm{B} (t)+ \mathbf{f}_\mathrm{B} (t)=0\, ,
\end{eqnarray}
\end{subequations}
where $\gamma$ is a drag coefficient and $\mathbf{f}_\mathrm{A} (t)$ and $\mathbf{f}_\mathrm{B} (t)$ are the Gaussian correlated stochastic forces accounting for thermal noise. These dynamics have been chosen as the simplest example to conceptualise the networking formalism such that additional effects such as drift and hydrodynamic interactions (see \textit{e.g.}~\cite{Doi&EdwardsBook, fredricksonCollectiveDynamicsPolymer1990}) have been deliberately omitted.
The Langevin equations can each be rewritten into a Martin-Siggia-Rose generating functional \cite{Martin:1973zz} as follows:
\begin{widetext}
\begin{subequations}
\label{eq:ZBM}
\begin{eqnarray}
Z_\mathrm{A}[\mathbf{J}_\mathrm{A}(t)] =\mathcal{N}\int [\mathrm{d} \mathbf{r}_\mathrm{A}(t)]  [\mathrm{d} \hat{\mathbf{r}}_\mathrm{A}(t)] [\mathrm{d} \mathbf{f}_\mathrm{A}(t)] \mathrm{e}^{\mathrm{i} \int_t \hat{\mathbf{r}}_\mathrm{A}(t)  \cdot \mathcal{L}_\mathrm{A}-\frac{1}{2 \lambda} \int_t |\mathbf{f}_\mathrm{A}(t) |^2+\mathrm{i}\int_t \mathbf{J}_\mathrm{A}(t) \cdot \mathbf{r}_\mathrm{A}(t)}\, ,
\label{eq:MSR-for-A}\\
Z_\mathrm{B}[\mathbf{J}_\mathrm{B}(t)] =\mathcal{N}\int [\mathrm{d} \mathbf{r}_\mathrm{B}(t)]  [\mathrm{d} \hat{\mathbf{r}}_\mathrm{B}(t)] [\mathrm{d} \mathbf{f}_\mathrm{B}(t)] \mathrm{e}^{\mathrm{i} \int_t \hat{\mathbf{r}}_\mathrm{B}(t)  \cdot \mathcal{L}_\mathrm{B}-\frac{1}{2 \lambda} \int_t |\mathbf{f}_\mathrm{B}(t) |^2+\mathrm{i} \int_t \mathbf{J}_\mathrm{B}(t) \cdot \mathbf{r}_\mathrm{B}(t)}\, 
\end{eqnarray}
\end{subequations}
\end{widetext}
where $\lambda  = 2 \gamma k_\mathrm{B} T$  according to the fluctuation-dissipation theorem. Here, the hatted variables are auxiliary variables which couple to the Langevin equations. Averages and correlation functions may be calculated by taking functional derivatives of the generating functionals with respect to their respective source terms $\mathbf{J}_\mathrm{A}$ and $\mathbf{J}_\mathrm{B}$ \textit{i.e.}
\begin{subequations}
\begin{eqnarray}
\langle \mathbf{r}_\mathrm{A}(t_1).... \mathbf{r}_\mathrm{A}(t_\nu) \rangle = \frac{1}{Z_\mathrm{A}[\mathbf{J}_\mathrm{A}(t)]}\left. \frac{\partial^\nu Z_\mathrm{A}[\mathbf{J}_\mathrm{A}(t)]}{\partial \mathbf{J}_\mathrm{A}(t_1)....\partial \mathbf{J}_\mathrm{A}(t_\nu)}    \right|_{\mathbf{J}_\mathrm{A}=0}\,, \\
\langle \mathbf{r}_\mathrm{B}(t_1).... \mathbf{r}_\mathrm{B}(t_\eta) \rangle = \frac{1}{Z_\mathrm{B}[\mathbf{J}_\mathrm{B}(t)]}\left. \frac{\partial^\eta Z_\mathrm{B}[\mathbf{J}_\mathrm{B}(t)]}{\partial \mathbf{J}_\mathrm{B}(t_1)....\partial \mathbf{J}_\mathrm{B}(t_\eta)}    \right|_{\mathbf{J}_\mathrm{B}=0}\,.
\label{eq:aveFormulas}
\end{eqnarray}
\end{subequations}
Incorporating  source terms that couple to the auxiliary variables allows the calculation of response functions with due care taken for causality (as reviewed in detail in \cite{Jouvet1979,Jensen1981a}). 
\subsubsection{Discrete interpretation}
We consider the first two terms in the exponent of eq.~\eqref{eq:SPnetworking}, since the third term will vanish in the upcoming approximations (see subsection \ref{subsec:densityFluctuations} ).  Recalling  eqs.~\eqref{eq:rhoB}--\eqref{eq:rhoA}, eq.~\eqref{eq:SPnetworking} can be rewritten as
\begin{eqnarray}
    \mathbb{Q}\approx \mathcal{N}_\Phi  \mathrm{e}^{\frac{1}{\tau}\int_{\mathbf{r},t} \sum_{m=1}^M \delta(\mathbf{r}-\mathbf{r_m}(t))\,\mathrm{ln}\left(1+\bar{\Phi}(\mathbf{r},t)\right)} \nonumber \\
    \times \mathrm{e}^{\frac{1}{\tau}\int_{\mathbf{r},t} \sum_{n=1}^N \delta(\mathbf{r}-\mathbf{r'_n}(t))\,\mathrm{ln}(\bar{\Phi}^*(\mathbf{r},t) )\,}\, .
\end{eqnarray}
Evaluating the spatial integrals and  replacing the temporal integrals with summations leads to
\begin{equation}
    \mathbb{Q}= \mathcal{N}_\Phi  \mathrm{e}^{ \sum_{t} \left( \sum_{m=1}^M \,\mathrm{ln}\left(1+\bar{\Phi}(\mathbf{r_m}(t),t)\right)+ \sum_{n=1}^N\,\mathrm{ln}(\bar{\Phi}^*(\mathbf{r'_n}(t),t) ) \right)}\,
\end{equation}
such that both terms can be brought out of the exponent to obtain
\begin{equation}
    \mathbb{Q} = \mathcal{N}_\Phi  \, \prod_{t} \prod_{\mathbf{r} \in \mathcal{A}} \prod_{\mathbf{r'} \in \mathcal{B}} q(\mathbf{r}_t,\mathbf{r}'_t,t) \,,
    \label{eq:Qdiscrete}
\end{equation}
where $q(\mathbf{r}_t,\mathbf{r}'_t,t) =\left(1+\bar{\Phi}(\mathbf{r},t)\right) \bar{\Phi}^*(\mathbf{r'},t)$\,.
Consider now, incorporating ~eq.~\eqref{eq:Qdiscrete} into a generating functional describing the dynamics of two separate sets of variables. Discretising the MSR generating functionals  given in eqs.~\eqref{eq:ZBM} and combining them using eq.~\eqref{eq:Qdiscrete} leads to
\begin{widetext}
    \begin{eqnarray}
        \mathbb{Z}[\mathbf{J}_{A,t},\mathbf{J}_{B,t}] =\mathcal{N}\prod_{t} \bigg\{ \prod_{\mathbf{r}_t \in \mathcal{A}}\int \mathrm{d} \mathbf{r}_t  \mathrm{d} \hat{\mathbf{r}}_t \mathrm{d} \mathbf{f}_{A,t} 
   \,\, \left(1+\bar{\Phi}(\mathbf{r}_t ,t)\right) \,\, \mathrm{e}^{\frac{\mathrm{i}}{\tau} \hat{\mathbf{r}}_t \cdot \mathcal{L}_\mathrm{A}[\mathbf{r}_t]-\frac{1}{2 \lambda \tau} |\mathbf{f}_{A,t} |^2 +\frac{\mathrm{i}}{\tau}  \mathbf{J}_{A,t} \cdot \mathbf{r}_t}\nonumber\\
   \prod_{\mathbf{r}'_t \in \mathcal{B}} \int  \mathrm{d} \mathbf{r}'_t  \mathrm{d} \hat{\mathbf{r}}'_t \mathrm{d} \mathbf{f}_{B,t}\, \,  \bar{\Phi}^*(\mathbf{r}'_t ,t) \mathrm{e}^{+\hat{\mathbf{r}}'_t \cdot \mathcal{L}_\mathrm{B}[\mathbf{r}'_t]-\frac{1}{2 \lambda \tau} |\mathbf{f}_{B,t} |^2 +\frac{\mathrm{i}}{\tau} \mathbf{J}_{B,t} \cdot \mathbf{r}'_t}\,\bigg\} \
   \label{eq:ZfullDiscretised} 
    \end{eqnarray}
\end{widetext}
where the shorthand notation $\mathbf{r}_t=\mathbf{r}_\mathrm{A}(t)$, $\mathbf{r}'_t=\mathbf{r}_\mathrm{B}(t)$, $\mathbf{f}_{A,t}=\mathbf{f}_\mathrm{A}(t)$, $\mathbf{f}_{B,t}=\mathbf{f}_\mathrm{B}(t)$, $\mathbf{J}_{A,t}=\mathbf{J}_\mathrm{A}(t)$ and $\mathbf{J}_{B,t}=\mathbf{J}_\mathrm{B}(t)$ has been used to indicate function values at each time $t$. Equation~\eqref{eq:ZfullDiscretised} gives, for each $\mathbf{r}_t \in \mathcal{A}$ and $\mathbf{r}'_t \in \mathcal{B}$, a product over all timesteps $t$, integrating over all possible values of $\mathbf{r}_t$ and $\mathbf{r}'_t$ weighted by the corresponding value of $q(\mathbf{r}_t,\mathbf{r}'_t,t)$ at each timestep . The discretised generating functional for the networked system of particles integrates trajectories of $\mathbf{r}_t$ and $\mathbf{r}'_t$ over all realisations of the stochastic forces $\mathbf{f}_\mathrm{A}(t)$ and $\mathbf{f}_\mathrm{B}(t)$ with the values of $q$ acting as weights in each step of each trajectory. If $\mathbf{r} \in \mathcal{A}$ and $\mathbf{r}' \in \mathcal{B}$, the possible values of the weights for each step in the trajectory are given by
\begin{align}
    \bar{\Phi}(\mathbf{r},t)= 
    \begin{cases}
        \tfrac{\rho_\mathrm{B}(\mathbf{r},t)}{(\rho_\mathrm{A}(\mathbf{r},t)-\rho_\mathrm{B}(\mathbf{r},t))} &\mathrm{if  } \,\mathbf{r} \in \mathcal{B}\\
        0 &\mathrm{if  } \,\mathbf{r} \notin \mathcal{B}
        \label{eq:phiDiscrete}
    \end{cases}
\end{align}
\begin{align}
    \bar{\Phi}^*(\mathbf{r}' ,t)= 
    \begin{cases}
         \tfrac{(\rho_\mathrm{A}(\mathbf{r'},t)-\rho_\mathrm{B}(\mathbf{r'},t))}{\alpha \tau}&\mathrm{if  }\, \mathbf{r}' \in \mathcal{A}\\
        0 &\mathrm{if  }\, \mathbf{r}' \notin \mathcal{A}
        \label{eq:phiStarDiscrete}
    \end{cases}
\end{align}

\begin{figure}
\begin{center}
\begin{tikzpicture}[scale = 0.8]
	\node [red] at (-4.5,1) {\large $t < \tau$: };
		\node [] (3) at (-1, 0.5) {};
		\node [] (4) at (-1.5, 1) {};
		\node [] (5) at (-0.5, 1) {};
		\node [] (21) at (-1, 0) {};
		\node [] (45) at (1, 0.5) {};
		\node [] (48) at (1, 0) {};
		\node [] (66) at (-2, 0) {};
		\node [] (67) at (8, 0) {};
		\node [] (68) at (1, 0.5) {};
		\node [] (69) at (0.5, 1) {};
		\node [] (70) at (1.5, 1) {};
		\node [] (71) at (3, 0.5) {};
		\node [] (72) at (3, 0) {};
		\node [] (73) at (3, 0.5) {};
		\node [] (74) at (2.5, 1) {};
		\node [] (75) at (3.5, 1) {};
		\node [] (76) at (5, 0.5) {};
		\node [] (77) at (5, 0) {};
		\node [] (78) at (5, 0.5) {};
		\node [] (81) at (7, 0.5) {};
		\node [] (82) at (7, 0) {};
		\node [] (83) at (7, 0.5) {};

		\node [red] at (-1.5,2) {\large $\mathbf{r}_1'(t)$};
		\draw  node[fill= red,circle,scale=2,above] (86) at (-1.5, 0.7) {};
		
		\node [red] at (1.3,2) {\large $\mathbf{r}_2'(t)$};
		\draw  node[fill= red,circle,scale=2,above]  at (1.3, 0.7) {};

		\draw [blue,dashed] (3.center) to (21.center) node[below]{\large $\mathbf{r}_1$};
		\draw [blue,dashed, bend right=90, looseness=1.75] (4.center) to (5.center);
		\draw [blue,dashed] (45.center) to (48.center) node[below]{\large $\mathbf{r}_2$};

		\draw [blue,dashed, bend right=90, looseness=1.75] (69.center) to (70.center);
		\draw [blue,dashed] (71.center) to (72.center) node[below]{\large $\mathbf{r}_3$};
		\draw [blue,dashed, bend right=90, looseness=1.75] (74.center) to (75.center);

\end{tikzpicture}

\begin{tikzpicture}[scale = 0.8]
	\node [red] at (-4.5,1) {\large $t = \tau$: };
		\node [] (3) at (-1, 0.5) {};
		\node [] (4) at (-1.5, 1) {};
		\node [] (5) at (-0.5, 1) {};
		\node [] (21) at (-1, 0) {};
		\node [] (45) at (1, 0.5) {};
		\node [] (48) at (1, 0) {};
		\node [] (66) at (-2, 0) {};
		\node [] (67) at (8, 0) {};
		\node [] (68) at (1, 0.5) {};
		\node [] (69) at (0.5, 1) {};
		\node [] (70) at (1.5, 1) {};
		\node [] (71) at (3, 0.5) {};
		\node [] (72) at (3, 0) {};
		\node [] (73) at (3, 0.5) {};
		\node [] (74) at (2.5, 1) {};
		\node [] (75) at (3.5, 1) {};
		\node [] (76) at (5, 0.5) {};
		\node [] (77) at (5, 0) {};
		\node [] (78) at (5, 0.5) {};
		\node [] (81) at (7, 0.5) {};
		\node [] (82) at (7, 0) {};
		\node [] (83) at (7, 0.5) {};

		\node [red] at (-1,2) {\large $\mathbf{r}_1'(t)$};
		\draw  node[fill= red,circle,scale=2,above] (86) at (-1, 0.7) {};
		
		\node [red] at (1,2) {\large $\mathbf{r}_2'(t)$};
		\draw  node[fill= red,circle,scale=2,above]  at (1, 0.7) {};

		\draw [blue,ultra thick] (3.center) to (21.center) node[below]{\large $\mathbf{r}_1$};
		\draw [blue,ultra thick, bend right=90, looseness=1.75] (4.center) to (5.center);
		\draw [blue,ultra thick] (45.center) to (48.center) node[below]{\large $\mathbf{r}_2$};

		\draw [blue,ultra thick, bend right=90, looseness=1.75] (69.center) to (70.center);
		\draw [blue,ultra thick] (71.center) to (72.center) node[below]{\large $\mathbf{r}_3$};
		\draw [blue,ultra thick, bend right=90, looseness=1.75] (74.center) to (75.center);

\end{tikzpicture}
\begin{tikzpicture}[scale = 0.8]
	\node [red] at (-4,1) {\large $\tau <t <2 \tau$: };
		\node [] (3) at (-1, 0.5) {};
		\node [] (4) at (-1.5, 1) {};
		\node [] (5) at (-0.5, 1) {};
		\node [] (21) at (-1, 0) {};
		\node [] (45) at (1, 0.5) {};
		\node [] (48) at (1, 0) {};
		\node [] (66) at (-2, 0) {};
		\node [] (67) at (8, 0) {};
		\node [] (68) at (1, 0.5) {};
		\node [] (69) at (0.5, 1) {};
		\node [] (70) at (1.5, 1) {};
		\node [] (71) at (3, 0.5) {};
		\node [] (72) at (3, 0) {};
		\node [] (73) at (3, 0.5) {};
		\node [] (74) at (2.5, 1) {};
		\node [] (75) at (3.5, 1) {};
		\node [] (76) at (5, 0.5) {};
		\node [] (77) at (5, 0) {};
		\node [] (78) at (5, 0.5) {};
		\node [] (81) at (7, 0.5) {};
		\node [] (82) at (7, 0) {};
		\node [] (83) at (7, 0.5) {};

		\node [red] at (-0.7,2) {\large $\mathbf{r}_1'(t)$};
		\draw  node[fill= red,circle,scale=2,above] (86) at (-0.7, 0.7) {};
		
		\node [red] at (0.4,2) {\large $\mathbf{r}_2'(t)$};
		\draw  node[fill= red,circle,scale=2,above]  at (0.4, 0.7) {};

		\draw [blue,dashed] (3.center) to (21.center) node[below]{\large $\mathbf{r}_1$};
		\draw [blue,dashed, bend right=90, looseness=1.75] (4.center) to (5.center);
		\draw [blue,dashed] (45.center) to (48.center) node[below]{\large $\mathbf{r}_2$};

		\draw [blue,dashed, bend right=90, looseness=1.75] (69.center) to (70.center);
		\draw [blue,dashed] (71.center) to (72.center) node[below]{\large $\mathbf{r}_3$};
		\draw [blue,dashed, bend right=90, looseness=1.75] (74.center) to (75.center);

\end{tikzpicture}
\begin{tikzpicture}[scale = 0.8]
	\node [red] at (-4.5,1) {\large $t =2 \tau$: };
		\node [] (3) at (-1, 0.5) {};
		\node [] (4) at (-1.5, 1) {};
		\node [] (5) at (-0.5, 1) {};
		\node [] (21) at (-1, 0) {};
		\node [] (45) at (1, 0.5) {};
		\node [] (48) at (1, 0) {};
		\node [] (66) at (-2, 0) {};
		\node [] (67) at (8, 0) {};
		\node [] (68) at (1, 0.5) {};
		\node [] (69) at (0.5, 1) {};
		\node [] (70) at (1.5, 1) {};
		\node [] (71) at (3, 0.5) {};
		\node [] (72) at (3, 0) {};
		\node [] (73) at (3, 0.5) {};
		\node [] (74) at (2.5, 1) {};
		\node [] (75) at (3.5, 1) {};
		\node [] (76) at (5, 0.5) {};
		\node [] (77) at (5, 0) {};
		\node [] (78) at (5, 0.5) {};
		\node [] (81) at (7, 0.5) {};
		\node [] (82) at (7, 0) {};
		\node [] (83) at (7, 0.5) {};

		\node [red] at (-1,2) {\large $\mathbf{r}_2'(t)$};
		\draw  node[fill= red,circle,scale=2,above] (86) at (-1, 0.7) {};
		
		\node [red] at (1,2) {\large $\mathbf{r}_1'(t)$};
		\draw  node[fill= red,circle,scale=2,above]  at (1, 0.7) {};

		\draw [blue,ultra thick] (3.center) to (21.center) node[below]{\large $\mathbf{r}_1$};
		\draw [blue,ultra thick, bend right=90, looseness=1.75] (4.center) to (5.center);
		\draw [blue,ultra thick] (45.center) to (48.center) node[below]{\large $\mathbf{r}_2$};

		\draw [blue,ultra thick, bend right=90, looseness=1.75] (69.center) to (70.center);
		\draw [blue,ultra thick] (71.center) to (72.center) node[below]{\large $\mathbf{r}_3$};
		\draw [blue,ultra thick, bend right=90, looseness=1.75] (74.center) to (75.center);

\end{tikzpicture}
\end{center}
    \caption{The dynamical networking functional  (eq.~\eqref{eq:SPnetworking} ) selects trajectories of the beads that satisfy the constraining nature of repeated cross-linking with the attachment points. Here the red circles correspond to the particles B and the blue cups to the particles A in eqs.~\eqref{eq:FbeforeSP}, \eqref{eq:SPsol1} and \eqref{eq:SPsol2}.}
    \label{fig:DynamicalNetworking}
\end{figure}
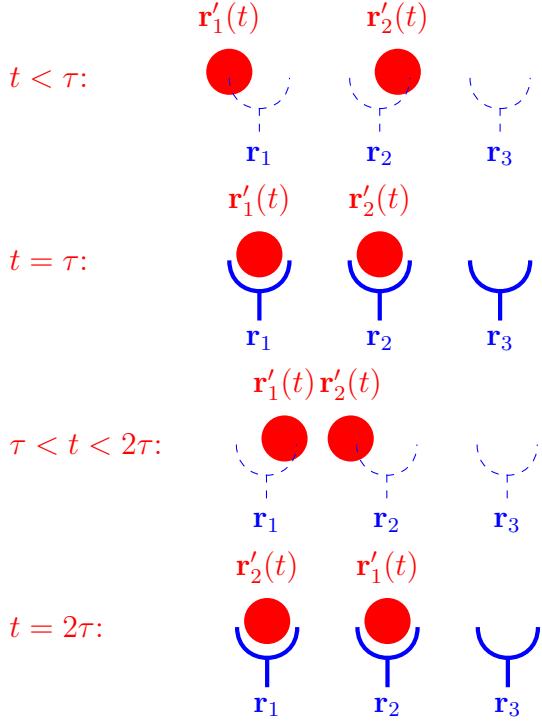

From eq.~\eqref{eq:phiDiscrete}, the  field $\bar{\Phi}$  diverges if the densities are exactly equal at a given time and position, but ( as discussed below eq.~\eqref{eq:effW_B} ) effectively models attachment sites that can only be networked to one bead at a time.  Figure~\ref{fig:DynamicalNetworking} depicts the scenario where 2 beads with positions $\mathbf{r}'_1(t)$ and $\mathbf{r}'_2(t)$ are networked to a set of 3 attachment sites with positions  $\mathbf{r}_1(t),\mathbf{r}_2(t)$ and $\mathbf{r}_3(t)$ with the corresponding generating functional given by
\begin{widetext}
 \begin{eqnarray}
        \mathbb{Z}[\mathbf{J}_{A,t},\mathbf{J}_{B,t}] =\mathcal{N}\prod_{t}  \int \mathrm{d} \mathbf{r}_{1,t}\,\mathrm{d} \mathbf{r}_{2,t} \,\mathrm{d} \mathbf{r}_{3,t} \mathrm{d} \mathbf{r}'_{1,t}\,\mathrm{d} \mathbf{r}'_{2,t} \,  \bigg\{\, \left(1+\bar{\Phi}(\mathbf{r}_{1,t} ,t)\right) \left(1+\bar{\Phi}(\mathbf{r}_{2,t} ,t)\right)\left(1+\bar{\Phi}(\mathbf{r}_{3,t} ,t)\right) \,\bar{\Phi}^*(\mathbf{r}'_{1,t} ,t)  \,\bar{\Phi}^*(\mathbf{r}'_{2,t} ,t) \nonumber\\
        \times\,\mathrm{e}^{\mathcal{F}_\mathrm{A}[\mathbf{r}_{1,t}]+\mathcal{F}_\mathrm{A}[\mathbf{r}_{2,t}]+\mathcal{F}_\mathrm{A}[\mathbf{r}_{3,t}]+\mathcal{F}_\mathrm{B}[\mathbf{r}'_{1,t}]+\mathcal{F}_\mathrm{B}[\mathbf{r}'_{2,t}]}\,\bigg\} \, .\nonumber\\
   \label{eq:Zfig2Discretised} 
    \end{eqnarray}
    \end{widetext}
    
If, for example, the product over $t$ is implemented at time intervals of $\tau$. Integrals over each of the positions of the beads and attachment points will be implemented at $t=\tau$,$t=2\tau$,$t=3\tau$ \textit{etc.} At each timestep, all possible values of each of the five positions are integrated over, weighting each possible value with a combination of the fields $\Phi$ and $\Phi^*$ at the corresponding time and position. Trajectories are formed by taking the product over multiple timesteps and therefore correspond to a product of the values of the weights. Trajectories that include a bead position that is not equal to that of an attachment point at any one of the discrete time steps $t=j \tau, j \in  \{0,1,2,3,..\}$, will be weighted by $\bar{\Phi}^*(\mathbf{r}' ,j \tau)=0$  and will therefore be discarded from the generating functional.  The remainder of the trajectories are assigned a sequence of finite weights given by $\bar{\Phi}^*(\mathbf{r}' ,j \tau)=\tfrac{(\rho_\mathrm{A}(\mathbf{r'},t)-\rho_\mathrm{B}(\mathbf{r'},t))}{\alpha \tau}$. This results in the \textit{hopping} behaviour of the beads, from attachment site to attachment site, at the discrete time intervals as depicted in Fig. \ref{fig:DynamicalNetworking}. The weights for the attachment sites are similar, with the additional term in $1+ \bar{\phi}$ allowing for attachment sites that are not networked to beads to be included in the generating functional. The attachment sites in Fig.~\ref{fig:DynamicalNetworking} are depicted as stationary, but in this example would exhibit diffusion behaviour similar to that of the beads. The networking formalism thus acts on the two dynamical systems, given separately by eqs.~\eqref{eq:ZBM}, by assigning weights to trajectories of the particles of both dynamical systems, such that only trajectories which satisfy the constraint in eq.~\eqref{eq:networkingConstraint} are selected as part of the generating function of the networked system.\\

\subsubsection{Collective description}
\label{subsec:densityFluctuations}
Moving towards collective descriptions of the dynamical system and implementing a Random Phase Approximation (RPA) as detailed in Appendix~\ref{app:RPA_BM}, one may obtain the following generating functionals for each set of particles
\begin{widetext} 
\begin{subequations}
    \label{eq:ZBMcollective}
    \begin{eqnarray}
    Z_\mathrm{A}[J_\mathrm{A}] =\mathcal{N}\int [\mathrm{d} \Delta \rho_\mathrm{A}]    \mathrm{e}^{-\frac{1}{2 } \int_{\mathbf{k}, \omega} \Delta \rho_\mathrm{A}(\mathbf{k}, \omega) \,S_{0,\mathrm{A}}^{-1}(\mathbf{k}, \omega)\, \Delta \rho_\mathrm{A}(-\mathbf{k}, -\omega)+\mathrm{i}\int_{\mathbf{k}, \omega}  J_\mathrm{A}(\mathbf{k}, \omega ) \,\Delta \rho_\mathrm{A}(-\mathbf{k}, -\omega)}\, ,\\
    Z_\mathrm{B}[J_\mathrm{B}] =\mathcal{N}\int [\mathrm{d} \Delta \rho_\mathrm{B}]    \mathrm{e}^{-\frac{1}{2 } \int_{\mathbf{k}, \omega} \Delta \rho_\mathrm{B}(\mathbf{k}, \omega) \,S_{0,\mathrm{B}}^{-1}(\mathbf{k}, \omega)\, \Delta \rho_\mathrm{B}(-\mathbf{k}, -\omega)+\mathrm{i}\int_{\mathbf{k}, \omega}  J_\mathrm{B}(\mathbf{k}, \omega ) \,\Delta \rho_\mathrm{B}(-\mathbf{k}, -\omega)}\, ,
    \end{eqnarray}
\end{subequations}
\end{widetext}
where the dynamic structure factors are given by 
\begin{subequations}
    \begin{eqnarray}
        S_{0,\mathrm{A}}(k, \omega) = \frac{D_\mathrm{A}k^2}{D_\mathrm{A}^2k^4 +\omega^2}\, ,\label{eq:structurefactorBMA}\\
        S_{0,\mathrm{B}}(k, \omega) = \frac{D_\mathrm{B}k^2}{D_\mathrm{B}^2k^4 +\omega^2}\,\, ,
        \label{eq:structurefactorBMB}
    \end{eqnarray}
\end{subequations}
and $D_\mathrm{A} =D_\mathrm{B} = \tfrac{\lambda}{2 \gamma^2}$ are the diffusion coefficients. The collective dynamics of both sets of particles can now be coupled to one another by combining both generating functionals \eqref{eq:ZBMcollective} along with the networking functional \eqref{eq:networkingAB} into a single generating functional as follows
\begin{multline}
\mathbb{Z}[J_\mathrm{A},J_\mathrm{B}] = \mathcal{N}\int [\mathrm{d} \Delta \rho_\mathrm{A}]  [\mathrm{d} \Delta \rho_\mathrm{B}]   \,\, \mathbb{Q}[\Delta \rho_\mathrm{A},\Delta \rho_\mathrm{B}] \,\\
    \times   \mathrm{e}^{-\frac{1}{2 } \int_{\mathbf{k}, \omega} \Delta \rho_\mathrm{A}(\mathbf{k}, \omega) \,S_{0,\mathrm{A}}^{-1}(\mathbf{k}, \omega)\, \Delta \rho_\mathrm{A}(-\mathbf{k}, -\omega)}\, \\
    \times \mathrm{e}^{-\frac{1}{2 } \int_{\mathbf{k}, \omega} \Delta \rho_\mathrm{B}(\mathbf{k}, \omega) \,S_{0,\mathrm{B}}^{-1}(\mathbf{k}, \omega)\, \Delta \rho_\mathrm{B}(-\mathbf{k}, -\omega)}\, \,\\
    \times   \mathrm{e}^{\mathrm{i}\int_{\mathbf{k}, \omega}  J_\mathrm{A}(\mathbf{k}, \omega ) \,\Delta \rho_\mathrm{A}(-\mathbf{k}, -\omega)+\mathrm{i}\int_{\mathbf{k}, \omega}  J_\mathrm{B}(\mathbf{k}, \omega ) \,\Delta \rho_\mathrm{B}(-\mathbf{k}, -\omega)}\, ,
    \label{eq:Zfull}
\end{multline}

Returning now to the networking functional  \eqref{eq:SPnetworking}, let $\rho_\mathrm{B}(\mathbf{r},t) = \bar{\rho}_\mathrm{B} + \Delta \rho_\mathrm{B} (\mathbf{r},t)$ and $\rho_\mathrm{A}(\mathbf{r},t) = \bar{\rho}_\mathrm{A} + \Delta \rho_\mathrm{A} (\mathbf{r},t)$, expanding the argument of the exponent of the networking functional up to second order in the fluctuations $\Delta \rho_\mathrm{A} (\mathbf{r},t)$ and  $\Delta \rho_\mathrm{B} (\mathbf{r},t)$ yields
\begin{eqnarray}
    \mathbb{Q}[\Delta \rho_\mathrm{B},\Delta \rho_\mathrm{A}]=\mathcal{N}_\Phi  \mathrm{e}^{- \frac{1}{\tau}\int_{\mathbf{r},t} \bar{\rho}_\mathrm{B} \, + \, \frac{1}{2\tau}\int_{\mathbf{r},t} \frac{(\Delta \rho_\mathrm{B} (\mathbf{r},t) - \Delta \rho_\mathrm{A} (\mathbf{r},t))^2}{\bar{\rho}_\mathrm{B}-\bar{\rho}_\mathrm{A}}} \nonumber \\ 
    \times \mathrm{e}^{\frac{1}{2\tau}\int_{\mathbf{r},t} \frac{\Delta \rho_\mathrm{A}^2 (\mathbf{r},t)}{\bar{\rho}_\mathrm{A}} } \, \, \, \, \, \, \, \, \, 
     \label{eq:networkingAB}
\end{eqnarray}

If we now consider the specific case where the attachment points remain fixed in time and space and are homogeneously distributed such that $\rho_\mathrm{A}(\mathbf{r},t) = \bar{\rho}_\mathrm{A}$, this becomes
\begin{equation}
    \mathbb{Q}[\Delta \rho_\mathrm{B},0]=\mathcal{N}_\Phi  \mathrm{e}^{ \frac{1}{2\tau}\int_{\mathbf{r},t} \int_{\mathbf{r}',t'}\frac{\Delta \rho_\mathrm{B} (\mathbf{r},t) \Delta \rho_\mathrm{B} (\mathbf{r}',t') }{\bar{\rho}_\mathrm{B}-\bar{\rho}_\mathrm{A}}\delta(\mathbf{r}-\mathbf{r}')\delta(t-t')}\, .
    \label{eq:networkingHomogenousAttPoints}
\end{equation}

In order to dynamically network the beads with collective density $\bar{\rho}_\mathrm{B} + \Delta \rho_\mathrm{B} (\mathbf{r},t)$  to attachment points with a  homogeneous, stationary density $\bar{\rho}_\mathrm{A}$, the networking functional \eqref{eq:networkingHomogenousAttPoints} can be incorporated into the generating functional \eqref{eq:Zfull} as follows 
\begin{multline}
    Z_{B,\mathbb{Q}} = \mathcal{N}\int[\mathrm{d} \, \Delta \rho_\mathrm{B}]  \, \,\mathbb{Q}[\Delta \rho_\mathrm{B},0]\\
    \times \mathrm{e}^{-\frac{1}{2}\int_{\mathbf{r},t} \int_{\mathbf{r}',t'} \Delta \rho_\mathrm{B} (\mathbf{r},t) S_{0,\mathrm{B}}^{-1} (\mathbf{r},\mathbf{r}',t,t') \Delta \rho_\mathrm{B} (\mathbf{r}',t')}\,.
\end{multline}
This leads to  the generating functional for the dynamics of the beads wherein only trajectories that satisfy the networking prescription are selected, \textit{i.e.}
\begin{widetext}
\begin{equation}
     Z_{B,\mathbb{Q}} = \mathcal{N}\int[\mathrm{d} \, \Delta \rho_\mathrm{B}]  \mathrm{e}^{ - \frac{1}{2}\int_{\mathbf{r},t} \int_{\mathbf{r}',t'} \Delta \rho_\mathrm{B} (\mathbf{r},t) (S_{0,\mathrm{B}}^{-1}(\mathbf{r}-\mathbf{r}',t-t')+W_\mathrm{B}(\mathbf{r}-\mathbf{r}',t-t') ) ( \Delta \rho_\mathrm{B} (\mathbf{r}',t')}
\end{equation}
\end{widetext}
where 
\begin{equation}
\begin{aligned}
   W_\mathrm{B}(\mathbf{r}-\mathbf{r}',t-t') &= w_\mathrm{B}\,\,\delta(\mathbf{r}-\mathbf{r}')\delta(t-t')\\
   &=\frac{\delta(\mathbf{r}-\mathbf{r}')\delta(t-t') }{\tau(\bar{\rho}_\mathrm{A}-\bar{\rho}_\mathrm{B})}\, 
   \label{eq:effW_B}
\end{aligned}
\end{equation}
may be interpreted as analogous to a \textit{time-dependent effective potential} (see \cite{Fantoni2011} for an equilibrium example within the networking formalism and \cite{rostiashviliLangevinDynamicsGlass1998} for a dynamical interpretation). This effective potential acts on the density fluctuations of the beads, \textit{i.e.}~$\Delta \rho_\mathrm{B}$, due to the dynamical networking with the attachment sites.  In this case, this time-dependent potential is short-ranged and repulsive (assuming $\bar{\rho}_\mathrm{A} > \bar{\rho}_\mathrm{B}$, which is true if the number of A sites exceeds the number of particles B), with a strength that decreases as $\tau$, the time interval between consecutive networking instances, increases. The form of the effective $W_\text{B}$ in eq.~\eqref{eq:effW_B} is a result of the networking constraint that permits only a single B particle to be paired with any A. When the densities of A and B are almost the same $\bar{\rho}_\mathrm{A} \gtrsim \bar{\rho}_\mathrm{B}$ the effective potential is strongly repulsive leading to much suppressed fluctuation correlations (if the positions of the A-particles remain fixed and homogeneously distributed).

Similar reasoning can be applied to obtain 
\begin{equation}
    \begin{aligned}
         W_\mathrm{A}(\mathbf{r}-\mathbf{r}',t-t') &= w_\mathrm{A}\,\,\delta(\mathbf{r}-\mathbf{r}')\delta(t-t')\\
         &= \frac{\bar{\rho}_\mathrm{B} }{\tau \bar{\rho}_\mathrm{A}(\bar{\rho}_\mathrm{A}-\bar{\rho}_\mathrm{B})}\delta(\mathbf{r}-\mathbf{r}')\delta(t-t')\, \, ,
        \label{eq:effW_A}
    \end{aligned}
\end{equation}
\textit{i.e.}~the time-dependent effective potential acting on the density fluctuations of the attachment points, \textit{i.e.}\ $\Delta \rho_\mathrm{A}$ due to the networking with the beads with density $\bar{\rho}_\mathrm{B}$. If one considers non-zero fluctuations in both $\rho_\mathrm{A}$ and $\rho_\mathrm{B}$ an additional effective potential, namely,
\begin{equation}
\begin{aligned}
   V_{AB}(\mathbf{r}-\mathbf{r}',t-t')&= v_\mathrm{AB}\,\,\delta(\mathbf{r}-\mathbf{r}')\delta(t-t')\\
   &= -\frac{\delta(\mathbf{r}-\mathbf{r}')\delta(t-t') }{\tau(\bar{\rho}_\mathrm{A}-\bar{\rho}_\mathrm{B})}\, 
   \label{eq:effV_AB}
\end{aligned}
\end{equation}
describes the short-ranged, attractive potential between the density fluctuations $\Delta \rho_\mathrm{A}$ and $\Delta \rho_\mathrm{B}$.
These \textit{effective potentials} (eqs.~\eqref{eq:effW_B}--\eqref{eq:effV_AB}) dictate the manner in which density variables will be coupled to one another dynamically when imposing the constraints via the networking functional \textit{i.e.}~eq.~\eqref{eq:networkingAB}. 

Note that these effective potentials only account for the effects of networking. Furthermore, the potentials are introduced into dynamical systems that also do not account for any interactions or excluded volume effects. Thus including these networking potentials into the system will result in a collapse of the system \cite{khokhlovSwellingCollapsePolymer1980}.  An additional repulsive potential $v$ can be included between particles of the same type to account for excluded volume effects \cite{Doi&EdwardsBook}. Replacing  $w_A$ with $w_A +v$ and $w_B$ with $w_B +v$ maintains the stability of the system if  the repulsive potential $v \geq  \tfrac{1}{2}\sqrt{w_\mathrm{A}^2 + 4 v_\mathrm{AB}^2 - 2 w_\mathrm{A} w_\mathrm{B} + w_\mathrm{B}^2}-\tfrac{1}{2}(w_\mathrm{A} +w_\mathrm{B})\,$.  This is discussed in more detail in Sec.~\ref{sec:polymerResults}.

\subsubsection{Final generating functional}
\label{subsubsec:BrownianExampleResults}

Returning to the example system with a full generating functional for networked Brownian particles, incorporating the effective potentials due to networking, as well as the repulsive potential $v$, leads to
\begin{eqnarray}
\mathbb{Z}[J_\mathrm{A},J_\mathrm{B}] = \mathcal{N}\int [\mathrm{d} \Delta \rho_\mathrm{A}]  [\mathrm{d} \Delta \rho_\mathrm{B}]    \mathrm{e}^{\mathrm{i}\int_{\mathbf{k}, \omega}  J_\mathrm{A}(\mathbf{k}, \omega ) \,\Delta \rho_\mathrm{A}(-\mathbf{k}, -\omega)}\nonumber \\ 
\times \mathrm{e}^{ \mathrm{i}\int_{\mathbf{k}, \omega}  J_\mathrm{B}(\mathbf{k}, \omega ) \,\Delta \rho_\mathrm{B}(-\mathbf{k}, -\omega)} \,\nonumber\\
    \times   \mathrm{e}^{-\frac{1}{2 } \int_{\mathbf{k}, \omega} \Delta \rho_\mathrm{A}(\mathbf{k}, \omega) \,(S_{0,\mathrm{A}}^{-1}(\mathbf{k}, \omega)+ w_\mathrm{A}+v)\, \Delta \rho_\mathrm{A}(-\mathbf{k}, -\omega)}\, \nonumber \\
    \times \mathrm{e}^{-\frac{1}{2 } \int_{\mathbf{k}, \omega} \Delta \rho_\mathrm{B}(\mathbf{k}, \omega) \,(S_{0,\mathrm{B}}^{-1}(\mathbf{k}, \omega)+ w_\mathrm{B}+v)\, \Delta \rho_\mathrm{B}(-\mathbf{k}, -\omega)}\, \, \nonumber\\
    \times 
    \mathrm{e}^{ -\int_{\mathbf{k}, \omega} \Delta \rho_\mathrm{A}(\mathbf{k}, \omega) \,v_{AB}\, \Delta \rho_\mathrm{B}(-\mathbf{k}, -\omega)}\,.\nonumber\\
    \label{eq:Zfull_Potentials}
\end{eqnarray}
The Gaussian path integrals over $\Delta \rho_\mathrm{A}$ and $\Delta \rho_\mathrm{B}$ can be implemented and the functional derivatives with respect to  $J_\mathrm{A}$ and $J_\mathrm{B}$ taken to obtain the correlation functions for the system of dynamically networked Brownian particles. Note that eq.~\eqref{eq:Zfull_Potentials} depends on the dynamic structure factors of both systems and the effective potentials due to networking, symbolically. Therefore, the correlation functions can be calculated in terms of symbolic expressions for the dynamic structure factors, allowing one to easily add networking into various dynamical systems by simply substituting the relevant dynamic structure factors into the correlation functions.  The correlation functions are given in Appendix\ref{app:corrFuns} as symbolic expressions of the dynamic structure factors and effective potentials, for ease of reference. 

Substituting the dynamic structure factors for the Brownian particles (eqs.~\eqref{eq:structurefactorBMA}--\eqref{eq:structurefactorBMB}) into the correlation function for $\Delta\rho_\mathrm{B}$ (\textit{i.e.}~eq.~\eqref{eq:BcorrNew}) this leads to the following correlation function

\begin{widetext}
   \begin{equation}
    \langle\!\langle\Delta \rho_\mathrm{B} (\mathbf{k}, \omega )  \Delta \rho_\mathrm{B} (-\mathbf{k}, -\omega )\rangle\!\rangle=\\
    \tfrac{D_\mathrm{B} k^2 \left(D_\mathrm{A} k^2 \left(D_\mathrm{A} k^2+w_\mathrm{B} +v\right)+\omega ^2\right)}{k^2 \omega ^2 \left(D_\mathrm{A} (D_\mathrm{A} k^2+w_\mathrm{A}+v)+D_\mathrm{B} \left(D_\mathrm{B} k^2+w_\mathrm{B}+v\right)\right)+D_\mathrm{A} D_\mathrm{B} k^4 \left(\left(D_\mathrm{A} k^2+w_\mathrm{A}+v\right) \left(D_\mathrm{B} k^2+w_\mathrm{B}+v\right)-v_\mathrm{AB}^2\right)+\omega ^4}\, ,
     \end{equation}
\end{widetext}

where $D_\mathrm{A}$and $D_\mathrm{B}$ are the original, \textit{non-networked} diffusion coefficients of the beads and attachment points, respectively.  The $\langle\!\langle ...\rangle\!\rangle$ notation is used to indicate that the average is taken over all realisations of both the stochastic forces $f_\mathrm{A}$ and  $f_\mathrm{B}$. The correlation function for the \textit{networked diffusing beads} depends not only on the correlation function of the \textit{non-networked diffusing beads}, but also that of the attachment points, the dynamical effective potentials (eqs.~\eqref{eq:effW_B}--\eqref{eq:effV_AB}) due to networking and the repulsive potential $v$. The networking potentials relevant to this example have been derived in Sec.~\ref{subsec:densityFluctuations} , but could also, should this be of interest, be substituted with alternative expressions derived from another  version of the networking functional (See \textit{e.g.}~\ref{sec:networkingAdv}). 

To further interpret this correlation function, consider once more the scenario where the attachment sites remain stationary such that $\rho_\mathrm{A}(\mathbf{r},t) = \bar{\rho}_\mathrm{A}$, then
 \begin{equation}
    \langle\Delta \rho_\mathrm{B} (\mathbf{k}, \omega )  \Delta \rho_\mathrm{B} (-\mathbf{k}, -\omega )\rangle=
    \tfrac{D_\mathrm{B} k^2}{\omega D_\mathrm{B}^2 k^4+D_\mathrm{B} (w_\mathrm{B}+v) k^2+\omega ^2}\, 
     \end{equation}
with corresponding poles
\begin{equation}
    \omega = \pm \mathrm{i}\sqrt{D_\mathrm{B} k^2 + w_\mathrm{B}+v}\sqrt{D_\mathrm{B}}\, k \,.
    \label{eq:BMpoles}
\end{equation}
For large $k$, $D_\mathrm{B} k^2 \gg w_\mathrm{B}+v$ such that this may be approximated as $\omega \approx \pm \mathrm{i} D_\mathrm{B} k^2$, which corresponds to purely diffusive dynamics with coefficient $D_\mathrm{B}$. Similarly, for small $k$, $\omega \approx \pm \mathrm{i} \sqrt{D_\mathrm{B}(w_\mathrm{B}+v)}\, k $, which is purely imaginary and linear in $k$ indicating that the dominant modes for the system are no longer purely diffusive, but rather overdamped modes which relax more slowly than diffusive modes. This suggests that at these length scales the effects of the intermittent networking of the beads to the attachment sites dominates over the diffusive motion of the beads. Upon further investigation of eq.~\eqref{eq:BMpoles}, a critical length scale $\ell_c = \sqrt{\tfrac{D_\mathrm{B}}{w_\mathrm{B}+v}}$ can be identified such that for lengths $\ell \ll \ell_c $ the beads diffuse with coefficient $D_\mathrm{B}$ , whilst for $\ell \gg \ell_c $ the beads exhibit collective dynamics at modes $\omega = \pm \frac{D_\mathrm{B}}{\ell_c} k$. Plugging in the effective networking potential (eq.~\eqref{eq:effW_B}) along with $v=v_\text{min}$, the critical length is given by $\ell_c = \tau (\bar{\rho}_\mathrm{A}-\bar{\rho}_\mathrm{B}) D_\mathrm{B} $. The length scale at which the collective behaviour becomes relevant thus depends on the diffusion coefficient of the beads, the timestep $\tau$ at which networking is implemented, as will as the average number of available attachment sites $\bar{\rho}_\mathrm{A}-\bar{\rho}_\mathrm{B}$.

This rudimentary example illustrates that the networking functional couples the dynamics of two sets of variables to one another and that the networking arises via a term analogous to a dynamical effective potential which may be read off directly from the networking functional. 

\subsubsection{Mechanical response}\label{Sec:Mechanical-Response}

The discussion has been based mainly on understanding density correlation functions in a Gaussian approximation for collective fields. However, the MSR formalism allows for a means of directly probing the response of the density fields to external fields to investigate the mechanical response of the networks to, for example, shear-inducing forces. Say, in the Brownian particle case, it is possible to add a position-dependent force field on particles A $\mathbf{F}_\text{ext}(\mathbf{r}_\text{A})$ to the Langevin equation for particles A in eq.~\eqref{eq:SPsol1}. This leads to the exponential factor in eq.~\eqref{eq:MSR-for-A} of the form
\begin{displaymath}
    \exp\left[   
        \mathrm{i} \int_t \hat{\mathbf{r}}_\text{A}\cdot\left( -\gamma \dot{\mathbf{r}}_\text{A}(t) +\mathbf{f}_\text{A}(t) 
        + \mathbf{F}_\text{ext}(\mathbf{r}_\text{A}(t)) 
        \right)
    \right]
\end{displaymath}
When written in terms of the collective coordinates $\hat{\rho}_\text{A}(\mathbf{r},t)$ (cf.~\eqref{eq:FH-collective-variables} and Appendix~\ref{app:RPA_BM}), this translates to a linear term in this collective variable of the form $\exp\left[\mathrm{i} \int_{r,t} \hat{\mathbf{\rho}}_\text{A}\cdot\mathbf{F}(\mathbf{r})\right]$. One can calculate linear response functions as shown in Appendix \ref{app:corrFuns} and discussed in Fredrickson and Helfand~\cite{fredricksonCollectiveDynamicsPolymer1990}, where the authors also show how this is understood for a coupled hydrodynamic field.

\section{Introducing a networking advantage}
\label{sec:networkingAdv}
The networking that has been implemented thus far, constrains the beads to the attachment points at consecutive discrete time steps, thereby requiring a fixed number of networking instances. Slight adaptations are introduced in the discussion below to relax these constraints such that the number of networking instances at each time step no longer remains fixed. This involves including an additional term to allow for beads to remain non-networked whilst assigning an advantage of $\mathrm{e}^\epsilon $ to beads that are networked at each time step. This results in a  weighted average in which the instances where beads are networked have a larger weight than that of non-networked beads. Introducing these adaptations into eq.~\eqref{eq:networkingConstraintRaw}, yields
\begin{widetext}
   \begin{equation}
   \mathbb{Q}=\int [ \mathrm{d}\Phi] [\mathrm{d} \Phi^*]\,\, \prod_j \left( \prod_m (1+\Phi(\mathbf{R_m},t_j))
    \, \prod_n (1 +\Phi^*(\mathbf{r_n},t_j)\mathrm{e}^\epsilon) \, \mathrm{e}^{-  \frac{1}{ \ell}\int_\mathbf{y} \, \Phi(\mathbf{y},t_j)\,\Phi^*(\mathbf{y},t_j)} \right)\, .
    \label{eq:networkingaAdvantage}    
    \end{equation} 
\end{widetext}
Again, the continuum limit can be considered and the collective variables may be introduced to obtain
\begin{equation}
 \mathbb{Q}[\rho_\mathrm{B}, \rho_\mathrm{A}]=\mathcal{N}_\Phi \int [ \mathrm{d}\Phi] [\mathrm{d} \Phi^*]\,\, \mathrm{e}^{ \mathbb{F}[\Phi,\Phi^*]} \,,
\label{eq:networkingConstraintAdv}    
\end{equation}
where
\begin{multline}
    \mathbb{F}[\Phi,\Phi^*] = \frac{1}{\tau}\int_{\mathbf{r},t} \rho_\mathrm{A}(\mathbf{r},t)\mathrm{ln}(1+\Phi(\mathbf{r},t)\\
    +\frac{1}{\tau}\int_{\mathbf{r},t} \rho_\mathrm{B}(\mathbf{r},t)\mathrm{ln}
\,(1+ \Phi^*(\mathbf{r},t) \mathrm{e}^\epsilon) \, -  \alpha \int_{\mathbf{y},t} \, \Phi(\mathbf{y},t)\,\Phi^*(\mathbf{y},t).
\end{multline}
 In the discussion to follow, it will be shown that  the networking advantage, $\epsilon$, may be utilised as a source term for a generating function to obtain an average for the number of networked beads.  This idea will then be utilised in order to identify the relevant physical saddle point solutions. 

\subsection{The average number of networked beads}
Taking the partial derivative of  ~eq.~\eqref{eq:networkingConstraintSP} with respect to the networking advantage $\epsilon$, we find
\begin{widetext}
    \begin{eqnarray}
        \frac{\partial \, \mathbb{Q}}{\partial \epsilon} = \int [ \mathrm{d}\Phi] [\mathrm{d} \Phi^*]\,\, \prod_j \left\{ \left( \prod_m (1+\Phi(\mathbf{R_m},t_j)) \right) \left(\mathrm{e}^\epsilon \mathcal{O}(\Phi^*) +2 \mathrm{e}^{2\epsilon} \mathcal{O}(\Phi^{*2}) + ... + N \mathrm{e}^{N\epsilon} \mathcal{O}(\Phi^{*N}) \right) \, \mathrm{e}^{-  \frac{1}{ \ell}\int_\mathbf{y} \, \Phi(\mathbf{y},t_j)\,\Phi^*(\mathbf{y},t_j)} \right\} \,.\nonumber\\ 
    \end{eqnarray}
\end{widetext}
When appropriately normalised, this yields a weighted average for the number of beads that have been networked, where the weights are determined by the networking advantage $\epsilon$, \textit{i.e.}
\begin{equation}
    \frac{1}{\mathbb{Q}}\frac{\partial \, \mathbb{Q}}{\partial \epsilon} = \frac{\partial 
 \, \mathrm{ln}\mathbb{Q}}{\partial \epsilon} = \langle \mathrm{number\, of\, networked\, beads}\rangle \, .
 \label{eq:numberNetworkedAve}
\end{equation}
Now applying ~eq.~\eqref{eq:numberNetworkedAve} to ~eq.~\eqref{eq:networkingConstraintAdv}, leads to

\begin{equation}
    \frac{\partial \, \mathrm{ln}\mathbb{Q}}{\partial \epsilon} = \int_{\mathbf{r},t} \frac{\rho_\mathrm{B}(\mathbf{r},t)\mathrm{e}^\epsilon}{\tau (1 + \mathrm{e}^\epsilon \Phi^* (\mathbf{r},t))}\Phi^* (\mathbf{r},t) \, .
    \label{eq:networkingAveForPhysSP}
\end{equation}

We can identify the saddle point equation, \textit{i.e.}~eq.~\eqref{eq:SP2-adv} in Sect.~\ref{sec:SPnetworkingAdv}, such that this can further be rewritten as
\begin{equation}
   \frac{\partial \, \mathrm{ln}\mathbb{Q}}{\partial \epsilon} = \alpha \int_{\mathbf{r},t} \bar{\Phi}(\mathbf{r},t) \bar{\Phi}^* (\mathbf{r},t)\, .
   \label{eq:numberNetworkedAvePhis}
\end{equation}

Thus $\alpha \bar{\Phi}(\mathbf{r},t) \bar{\Phi}^* (\mathbf{r},t)$ corresponds to the number density of networked beads at position $\mathbf{r}$ and time $t$. For further comments on the use of the saddle point solutions of the fields $\Phi$ and $\Phi^*$ in this context, refer to Appendix~\ref{app:SPforAverage}.

\subsection{The saddle point approximation}
\label{sec:SPnetworkingAdv}

The saddle point solutions $\bar{\Phi}$ and $\bar{\Phi}^*$ are given by :
\begin{subequations}
\begin{eqnarray}
0=\left.\frac{\partial \mathbb{F}}{\partial \Phi(r,t)} \right|_{\bar{\Phi}^* ,\bar{\Phi}}= -\alpha \bar{\Phi}^*(\mathbf{r},t) + \tfrac{\rho_\mathrm{A}(\mathbf{r},t)}{\tau (1+\bar{\Phi}(\mathbf{r},t))} \label{eq:SP1-adv}\, \, \, \, \,  \\
0=\left.\frac{\partial \mathbb{F}}{\partial \Phi^*(r,t)} \right|_{\bar{\Phi}^* ,\bar{\Phi}}= -\alpha \bar{\Phi}(\mathbf{r},t) + \tfrac{\rho_\mathrm{B}(\mathbf{r},t)\mathrm{e}^\epsilon}{\tau (1+\bar{\Phi}^*(\mathbf{r},t)\mathrm{e}^\epsilon)} \label{eq:SP2-adv}\, \, \, \, \, 
\end{eqnarray}
\end{subequations}
Solving the simultaneous eqs.(\ref{eq:SP1-adv}) and (\ref{eq:SP2-adv}) 
yields two solutions for $\bar{\Phi}$ and $\bar{\Phi}^*$ each, such that the relevant physical solutions need to be identified. The average number of networked particles, \textit{i.e.}~eq.~\eqref{eq:networkingAveForPhysSP}, must provide a physical quantity which is real and positive. Substituting both possible solutions into eq.~\eqref{eq:networkingAveForPhysSP}, shows that both of the solutions yield a positive and real expression under different criteria. \\

The solutions given by
\begin{subequations}
\label{eq:NetAdvSPsols}
\begin{multline}
\bar{\Phi}^*(\mathbf{r},t) =\frac{1}{2\alpha} \left[- \alpha \mathrm{e}^{-\epsilon} + \frac{1}{\tau}\left(\rho_\mathrm{A}(\mathbf{r},t)-\rho_\mathrm{B}(\mathbf{r},t)\right)\right.\\
\left. -\mathrm{e}^{-\epsilon} \sqrt{(\alpha +\frac{\mathrm{e}^\epsilon}{\tau}\left(\rho_\mathrm{B}(\mathbf{r},t)-\rho_\mathrm{A}(\mathbf{r},t)\right)^2+ \frac{4 \alpha \mathrm{e}^\epsilon }{\tau}\rho_\mathrm{A}(\mathbf{r},t)}  \right]
\end{multline}
    \begin{equation}
        \bar{\Phi}(\mathbf{r},t) =\mathrm{e}^{\epsilon}\left( \bar{\Phi}^*(\mathbf{r},t)  - \frac{ \rho_\mathrm{A}(\mathbf{r},t) -\rho_\mathrm{B}(\mathbf{r},t)}{\alpha \tau}\right)
\end{equation}
\end{subequations}
lead to positive and real values for ~eq.~\eqref{eq:numberNetworkedAvePhis} for all non-negative values of $\rho_\mathrm{A}$ and $\rho_\mathrm{B}$. The other set of solutions for $ \bar{\Phi}$ and $ \bar{\Phi}^*$ leads to a negative value for  ~eq.~\eqref{eq:numberNetworkedAvePhis}, \textit{i.e.}~the number of networked beads, when $\rho_\mathrm{A} =0$, which is unphysical. This leads to the conclusion that eqs.~\eqref{eq:NetAdvSPsols} are the physical solutions to $ \bar{\Phi}$ and $ \bar{\Phi}^*$. This is also confirmed by the results obtained in Sect.~\ref{sec:polymerResults} and Fig.~\ref{fig:numNetworkedPlotswithV}.\\

Substituting eqs.~\eqref{eq:NetAdvSPsols} into  
\begin{equation}
 \mathbb{Q}_\mathrm{SP}[\rho_\mathrm{B}, \rho_\mathrm{A}]=\mathcal{N}_\Phi \mathrm{e}^{ \mathbb{F}[\bar{\Phi},\bar{\Phi}^*]} \,.
\label{eq:networkingConstraintSP}    
\end{equation}
amounts to the saddle point approximation. In addition to this approximation, small fluctuation expansions can be implemented for the collective variables as before, to obtain
\begin{multline}
    \mathbb{Q}[\Delta \rho_\mathrm{B}, \Delta \rho_\mathrm{A}] = \mathcal{N}\mathrm{e}^{\frac{1}{\tau}\int_{\mathbf{r},t}\Delta \rho_\mathrm{A}(\mathbf{r},t) \mathrm{ln}\left(\frac{\mathrm{e}^\epsilon (\bar{\rho}_\mathrm{B} - \bar{\rho}_\mathrm{A}) + \tau(\alpha + \eta[\bar{\rho}_\mathrm{B}, \bar{\rho}_\mathrm{A}])}{2 \alpha \tau}\right)} \\
    \times \mathrm{e}^{ +\frac{1}{\tau}\int_{\mathbf{r},t}\Delta \rho_\mathrm{B}(\mathbf{r},t) \mathrm{ln}\left(\frac{\mathrm{e}^\epsilon (\bar{\rho}_\mathrm{A} - \bar{\rho}_\mathrm{B}) + \tau(\alpha + \eta[\bar{\rho}_\mathrm{B}, \bar{\rho}_\mathrm{A}])}{2 \alpha \tau} \right)}\\ \times \mathrm{e}^{\frac{\mathrm{e}^\epsilon}{\tau^2 \eta[\bar{\rho}_\mathrm{B}, \bar{\rho}_\mathrm{A}]}\int_{\mathbf{r},t}\Delta \rho_\mathrm{B}(\mathbf{r},t)\,\Delta \rho_\mathrm{A}(\mathbf{r},t)-\frac{\mathrm{e}^{2\epsilon}\zeta[\bar{\rho}_\mathrm{B}, \bar{\rho}_\mathrm{A}]}{\tau} \int_{\mathbf{r},t}\bar{\rho}_\mathrm{B}\,\Delta \rho_\mathrm{A}^2(\mathbf{r},t)} \\
    \times \mathrm{e}^{-\frac{\mathrm{e}^{2\epsilon}\zeta[\bar{\rho}_\mathrm{B}, \bar{\rho}_\mathrm{A}]}{\tau }\int_{\mathbf{r},t}\bar{\rho}_\mathrm{A} \,\Delta \rho_\mathrm{B}^2(\mathbf{r},t)}  
\end{multline}
where
\begin{equation}
    \eta[\bar{\rho}_\mathrm{B}, \bar{\rho}_\mathrm{A}] =\sqrt{\alpha^2 + \frac{1}{\tau^2}(\bar{\rho}_\mathrm{B}-\bar{\rho}_\mathrm{A})^2\mathrm{e}^{2\epsilon} + \frac{2 \alpha}{\tau}(\bar{\rho}_\mathrm{B}+\bar{\rho}_\mathrm{A})\mathrm{e}^{\epsilon}}
\end{equation}
and 
\begin{widetext}
\begin{equation}
       \zeta[\bar{\rho}_\mathrm{B}, \bar{\rho}_\mathrm{A}] = \frac{\alpha  \tau ^2 (\alpha +\eta[\bar{\rho}_\mathrm{B}, \bar{\rho}_\mathrm{A}] )+\tau  e^{\epsilon } (2 \alpha +\eta[\bar{\rho}_\mathrm{B}, \bar{\rho}_\mathrm{A}] ) (\bar{\rho}_\mathrm{A}+\bar{\rho}_\mathrm{B})+e^{2 \epsilon } (\bar{\rho}_\mathrm{A}-\bar{\rho}_\mathrm{B})^2}{\left(\alpha ^2 \tau ^2+2 \alpha  \tau  e^{\epsilon } (\bar{\rho}_\mathrm{A}+\bar{\rho}_\mathrm{B})+e^{2 \epsilon } (\bar{\rho}_\mathrm{A}-\bar{\rho}_\mathrm{B})^2\right) \left(\tau  (\alpha +\eta[\bar{\rho}_\mathrm{B}, \bar{\rho}_\mathrm{A}] )+e^{\epsilon } (\bar{\rho}_\mathrm{A}+\bar{\rho}_\mathrm{B})\right)^2}
\end{equation}
\end{widetext}
From this the effective potentials due to networking can again be determined \textit{i.e.}
\begin{eqnarray}
    W_\mathrm{B}(\mathbf{r}-\mathbf{r}',t-t') = w_\mathrm{B}\, \delta(\mathbf{r}-\mathbf{r}')\delta(t-t')\, , \, \, \, \, \, \,\, \, \, \, \, \, \, \, \label{eq:WBAdv}\\
     W_\mathrm{A}(\mathbf{r}-\mathbf{r}',t-t') = w_\mathrm{A}\,\delta(\mathbf{r}-\mathbf{r}')\delta(t-t')\,  , \, \, \, \, \, \,\, \, \, \, \, \, \, \, \label{eq:WAAdv}\\
     V_\mathrm{AB}(\mathbf{r}-\mathbf{r}',t-t') = v_\mathrm{AB}\,\delta(\mathbf{r}-\mathbf{r}')\delta(t-t') \, ,\, \, \, \, \, \, \, \,\, \, \, \, \, \, \, \,
     \label{eq:V_ABAdv}
\end{eqnarray}
where
\begin{eqnarray}
    w_\mathrm{B}= \tfrac{2\bar{\rho}_\mathrm{A} }{\tau }\mathrm{e}^{2\epsilon}\zeta[\bar{\rho}_\mathrm{B}, \bar{\rho}_\mathrm{A}] \, , \, \, \, \, \, \,\, \, \, \, \, \, \, \, \label{eq:wBAdv}\\
     w_\mathrm{A} =\tfrac{2\bar{\rho}_\mathrm{B} }{\tau }\mathrm{e}^{2\epsilon}\zeta[\bar{\rho}_\mathrm{B}, \bar{\rho}_\mathrm{A}] \,  , \, \, \, \, \, \,\, \, \, \, \, \, \, \, \label{eq:wAdv}\\
     v_\mathrm{AB} = -\tfrac{\mathrm{e}^{\epsilon}}{\tau^2 \eta[\bar{\rho}_\mathrm{B}, \bar{\rho}_\mathrm{A}]} \, ,\, \, \, \, \, \, \, \,\, \, \, \, \, \, \, \,
     \label{eq:v_ABAdv}
\end{eqnarray}
which are again short-ranged as expected. If we consider the case where $\bar{\rho}_\mathrm{A} = \bar{\rho}_\mathrm{B}$, eqs.~\eqref{eq:WBAdv}--\eqref{eq:WAAdv} correspond to repulsive interaction potentials whilst eq.~\eqref{eq:V_ABAdv} corresponds to an attractive interaction potential.
\section{Intermittently cross-linked polymer chains} 
Consider two solutions of polymers with densities $\rho_\mathrm{A}$ and $\rho_\mathrm{B}$, respectively. The networking functional from Section \ref{sec:networkingAdv} can be used to model a mixture where each of the polymer solutions contains a different type of polymer such that cross-linking is only permitted between polymer segments of species A and species B. 
\subsection{Polymer dynamics}
From Ref. \cite{fredricksonCollectiveDynamicsPolymer1990}, 
the dynamic structure factors for the non-networked solutions of polymer chains in the scenario where $k R_\mathrm{g} \ll 1$ are given by

\begin{subequations}
     \begin{eqnarray}
     \label{eq:Spolymers}
        S_{0,\mathrm{A}}(k, \omega) =  \frac{2 \gamma_\mathrm{A}  k^2}{\gamma_\mathrm{A}^2 \omega ^2+L_\mathrm{A}^ {-2} k^4 } ,\\
        S_{0,\mathrm{B}}(k, \omega) = \frac{2 \gamma_\mathrm{B}  k^2}{\gamma_\mathrm{B}^2 \omega ^2+L_\mathrm{B}^ {-2} k^4 } 
    \end{eqnarray}
\end{subequations}
and the linear response functions by
\begin{subequations}
    \begin{eqnarray}
     \label{eq:Chipolymers}
        \chi_{0,\mathrm{A}}(k, \omega) =  \frac{2\gamma_\mathrm{A}  k^2}{\gamma_\mathrm{A} \omega + \mathrm{i} L_\mathrm{A}^ {-1} k^2 } ,\\
        \chi_{0,\mathrm{B}}(k, \omega) = \frac{ \gamma_\mathrm{B}  k^2}{\gamma_\mathrm{B} \omega + \mathrm{i} L_\mathrm{B}^ {-1} k^2} \,.
    \end{eqnarray}
\end{subequations}
Here $\gamma_\mathrm{A}$ and $\gamma_\mathrm{B}$ are the drag coefficients whilst $L_\mathrm{A}$ and $L_\mathrm{B}$ are the lengths of the polymers in each solution. The poles of these dynamic structure factors and response functions are given by
\begin{subequations}
    \begin{equation}
        \omega = \pm \frac{\mathrm{i} k^2}{\gamma_\mathrm{A}L_\mathrm{A}}
        \label{eq:polymerApoles}
    \end{equation}
    \begin{equation}
        \omega = \pm \frac{\mathrm{i} k^2}{\gamma_\mathrm{B}L_\mathrm{B}}
        \label{eq:polymerBpoles}
    \end{equation}
\end{subequations}
indicating purely diffusive behaviour with diffusion coefficients inversely proportional to the length of the polymers and drag coefficients in each solution.
\subsection{Results}
\label{sec:polymerResults}
 Dynamic structure factors  and linear response functions for a polymer mixture with cross-linking introduced via the proposed networking formalism are presented and analysed in the discussion to follow. Evidently, upon the addition of networking the system no longer presents purely diffusive behaviour and shows a less prominent diffusive contribution as the networking advantage $\epsilon$ is increased.  Introducing networking to create cross-linking between the polymers of species A and species B in this manner introduces an attractive potential between the polymers which is known to cause a collapse of the system \cite{khokhlovSwellingCollapsePolymer1980}. The discussion below concludes by confirming  that within the current formalism the system does collapse --as expected-- and therefore necessitates the inclusion of an additional repulsive potential \cite{Doi&EdwardsBook} between polymers of the same species to ensure stability. 

\begin{figure*}[htbp]
    \centering
    \subfigure[]{
        \includegraphics[width=0.45\textwidth]{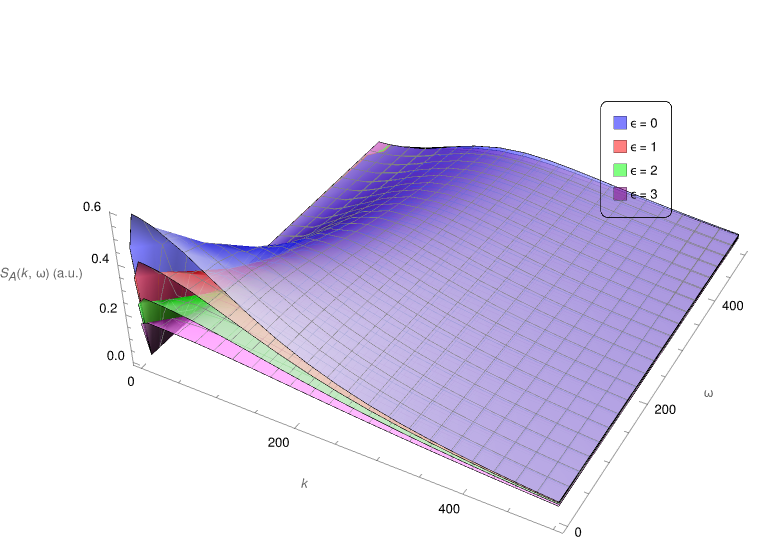}
        \label{fig:SAplotwithV}
    }
    \subfigure[]{
        \includegraphics[width=0.45\textwidth]{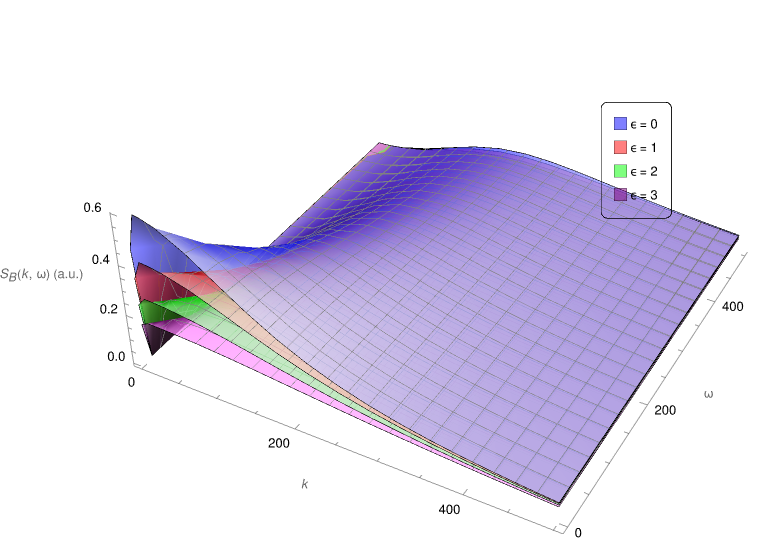}
        \label{fig:SBplotwithV}
    }

    \caption{Dynamic structure factors for the polymers with density $\bar{\rho}_\mathrm{A} + \Delta \rho_\mathrm{A} (k, \omega)$ and $\bar{\rho}_\mathrm{B} + \Delta \rho_\mathrm{B} (k, \omega)$, respectively, in arbitrary units for the cross-linked polymer mixture with repulsive effective potential $v=v_\mathrm{min}$ (see ~eq.~\eqref{eq:Vmin}) between polymers of the same type. Here $\bar{\rho}_\mathrm{A} = \bar{\rho}_\mathrm{B} = 1$,$\gamma_\mathrm{A} = \gamma_\mathrm{B} =1$, $L_\mathrm{A}=L_\mathrm{B}=100$ with varying values of $\epsilon$. }
    \label{fig:NetworkedStructureFactorPlotswithV}
\end{figure*}

Substituting eqs.~\eqref{eq:Spolymers} along with  the effective potentials due to networking eqs.~\eqref{eq:WBAdv}--\eqref{eq:V_ABAdv} into eqs.~\eqref{eq:AcorrNew}--\eqref{eq:ABcorrNew} gives the correlation functions for the networked system, \textit{i.e.}\ the polymer mixture with cross-linking occurring  between polymers with densities $\bar{\rho}_\mathrm{A} + \Delta \rho_\mathrm{A} (k, \omega)$ and $\bar{\rho}_\mathrm{B} + \Delta \rho_\mathrm{B} (k, \omega)$. The structure factors in Fig.~\ref{fig:SAplotwithV}-\ref{fig:SBplotwithV}  depict a peak at low $k$ and $\omega$ values, which drops off for higher values as expected for a dynamical structure factor. These peaks occur at the poles of the structure factors,
\begin{widetext}
 \begin{equation}
 \omega = 
\resizebox{\textwidth}{!}{$
   \pm \sqrt{\tfrac{- \left(\gamma_\mathrm{A}^2 L_\mathrm{A}^2+\gamma_\mathrm{B}^2 L_\mathrm{B}^2\right)k^4+2 \gamma_\mathrm{A} \gamma_\mathrm{B} L_\mathrm{A}^2 L_\mathrm{B}^2 (\gamma_\mathrm{B} w_\mathrm{A}+\gamma_\mathrm{A} w_\mathrm{B})k^2 \pm\sqrt{\left(\gamma_\mathrm{A}^2 L_\mathrm{A}^2-\gamma_\mathrm{B}^2 L_\mathrm{B}^2\right)^2 k^8 +4 \gamma_\mathrm{A} \gamma_\mathrm{B} k^6 L_\mathrm{A}^2 L_\mathrm{B}^2 (\gamma_\mathrm{A} L_\mathrm{A}+\gamma_\mathrm{B} L_\mathrm{B}) (\gamma_\mathrm{A} L_\mathrm{A}-\gamma_\mathrm{B} L_\mathrm{B}) (\gamma_\mathrm{A} w_\mathrm{B}-\gamma_\mathrm{B} w_\mathrm{A})k^4+4 \gamma_\mathrm{A}^2 \gamma_\mathrm{B}^2 L_\mathrm{A}^4 L_\mathrm{B}^4 \left(\gamma_\mathrm{B}^2 w_\mathrm{A}^2-2 \gamma_\mathrm{A} \gamma_\mathrm{B} w_\mathrm{A} w_\mathrm{B}+4 \gamma_\mathrm{A} \gamma_\mathrm{B} v_\mathrm{AB}^2+\gamma_\mathrm{A}^2 w_\mathrm{B}^2\right)}}{2\gamma_\mathrm{A}^2 \gamma_\mathrm{B}^2 L_\mathrm{A}^2 L_\mathrm{B}^2}} \,,
    \label{eq:divergenceFull}
        $}
\end{equation}   
\end{widetext}
which can be approximated as
\begin{widetext}
 \begin{equation}
    \omega = \pm\sqrt{\tfrac{-((w_\mathrm{A}+v)\gamma_\mathrm{A}+(w_\mathrm{B}+v)\gamma_\mathrm{B})\pm\sqrt{4 \gamma_\mathrm{A} \gamma_\mathrm{B} v_{AB}^2-4 \gamma_\mathrm{A} \gamma_\mathrm{B}( w_\mathrm{A}+v)( w_\mathrm{B}+v)+\left((w_\mathrm{A}+v)\gamma_\mathrm{A}+(w_\mathrm{B}+v)\gamma_\mathrm{B}\right)^2}}{\gamma_\mathrm{A} \gamma_\mathrm{B}}}\, k +  \mathcal{O}(k^3)\,.
    \label{eq:SpolesWithV}
\end{equation}   
\end{widetext}
for small $k$. Including the networking and repulsive potentials therefore shifts the system from purely diffusive behaviour as seen in the poles of the non-networked dynamic structure factors eqs.~\eqref{eq:polymerApoles}-\eqref{eq:polymerBpoles} to poles with a more complicated $k$ -dependence. For large wavelengths (small $k$) the largest contribution is now a linear term in $k$, which is characteristic of waves propagating with a linear dispersion in systems with elasticity (see \textit{e.g.}~\cite{chaikinPrinciplesCondensedMatter2013}). The coefficient of this linear term in $k$ is a combination of the networking potentials and viscous drag coefficients of the polymers in solution and depending on the signs results in both real and imaginary poles. There is also a higher order correction, but there is no $k^2$ contribution corresponding to diffusive behaviour as was present in both dynamic structure factors of the non-networked polymer solutions. Thus, introducing the networking and repulsive potentials has transitioned the system away from purely diffusive behaviour, as expected. Whilst the diffusive behaviour is likely still present, its effect has been diminished and is combined with that of the effective potentials in both the linear and higher order $k$ contributions. This is also evident in Fig.~\ref{fig:SAplotwithV}-\ref{fig:SBplotwithV} , since the peaks of the dynamic structure factors flatten as the networking advantage $\epsilon$ is increased, \textit{i.e.}\ as the likelihood or strength of cross-links is increased the diffusive behaviour corresponding to the peak becomes less prominent. 
The linear response functions are obtained by substituting eqs.~\eqref{eq:Chipolymers} along with  the effective potentials due to networking eqs.~\eqref{eq:WBAdv}--\eqref{eq:V_ABAdv} into eqs.~\eqref{eq:AresponseNew}--\eqref{eq:BAresponseNew}. Since the general form of these response functions are similar and they all share the same poles as the dynamic structure factors, only the response function for polymer $A$  will be discussed in more detail here. In the long time limit this is given by
\begin{widetext}
 \begin{equation}
    \langle\!\langle\,  \Delta \rho_\mathrm{A} ( \mathbf{k}, \omega) \Delta \hat{\rho}_\mathrm{A} ( -\mathbf{k},- \omega)\,\rangle\!\rangle _{\omega\to0}=\frac{-\mathrm{i} \gamma_\mathrm{A} k^2 L_\mathrm{A}\left(k^2+2 \gamma_\mathrm{B} L_\mathrm{B}^2 (v+w_\mathrm{B})\right)}{k^4+2 k^2 \left(\gamma_\mathrm{A} L_\mathrm{A}^2 (v+w_\mathrm{A})+\gamma_\mathrm{B} L_\mathrm{B}^2 (v+w_\mathrm{B})\right)+4 \gamma_\mathrm{A} \gamma_\mathrm{B} L_\mathrm{A}^2 L_\mathrm{B}^2 \left((v+w_\mathrm{A}) (v+w_\mathrm{B})-v_\mathrm{AB}^2\right)}\, ,
    \label{eq:responseAlongTime}
\end{equation}
\end{widetext}
which in the absence of the networking potentials gives $ -\mathrm{i} \gamma_\mathrm{A} L_\mathrm{A}$, consistent with the response function of the non-networked polymer solution (see eq.~\eqref{eq:Chipolymers}).   For small length scales (or large $k$), eq.~\eqref{eq:responseAlongTime}  becomes ,$\langle\!\langle\,  \Delta \rho_\mathrm{A} ( \mathbf{k}, \omega) \Delta \hat{\rho}_\mathrm{A} ( -\mathbf{k},- \omega)\,\rangle\!\rangle _{k\to\infty}= -\mathrm{i} \gamma_\mathrm{A} L_\mathrm{A}$, whilst for large length scales (or small $k$), $\langle\!\langle\,  \Delta \rho_\mathrm{A} ( \mathbf{k}, \omega) \Delta \hat{\rho}_\mathrm{A} ( -\mathbf{k},- \omega)\,\rangle\!\rangle _{k\to0}= 0$, indicating that there is no response to fields that are spatially uniform as one also finds for the non- networked polymer solution $\mathrm{A}$. Thus, the long time response function for polymer solution $\mathrm{A}$ only deviates from that of the non-networked system on intermediate length scales where it exhibits a more intricate $k$-dependence  related to the effective potentials due to networking as well as the parameters of both polymer solutions $\mathrm{A}$ and $\mathrm{B}$, as given in eq.~\eqref{eq:responseAlongTime}.

\begin{figure}[htbp]
    \centering
        \includegraphics[width=0.5\textwidth]{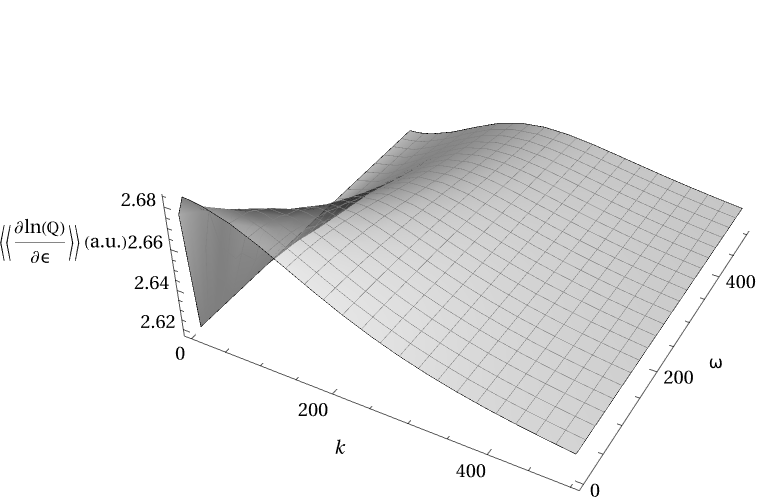}
    \caption{The average number density of networked particles or cross-links due to dynamical networking in the polymer mixture with repulsive effective potential $v=v_\mathrm{min}$ (see ~eq.~\eqref{eq:Vmin}) between polymers of the same type. Here $\bar{\rho}_\mathrm{A} = \bar{\rho}_\mathrm{B} = 1$,$\gamma_\mathrm{A} = \gamma_\mathrm{B} =1$, $L_\mathrm{A}=L_\mathrm{B}=100$, $\epsilon=0$.}
    \label{fig:numNetworkedPlotswithV}
\end{figure}

In addition to dynamic structure factors  and linear response functions for the networked system, the average number of cross-links or instances of networking can be determined. Recalling ~eq.~\eqref{eq:numberNetworkedAvePhis}, the number density of networked polymer beads is given by  $\alpha \bar{\Phi}(\mathbf{r},t) \bar{\Phi}^* (\mathbf{r},t)$, take the spatial and temporal Fourier transforms and once again let $\rho_\mathrm{A} (k, \omega)=\bar{\rho}_\mathrm{A} + \Delta \rho_\mathrm{A} (k, \omega)$ and $\rho_\mathrm{B} (k, \omega)=\bar{\rho}_\mathrm{B} + \Delta \rho_\mathrm{B} (k, \omega)$ and expand up to second order in $\Delta \rho_\mathrm{A} (k, \omega)$ and $ \Delta \rho_\mathrm{B} (k, \omega)$. Taking the average of this yields 
\begin{widetext}
\begin{eqnarray}
   \langle\!\langle \alpha \bar{\Phi}(k, \omega) \bar{\Phi}^*(k, \omega)\rangle\!\rangle = \frac{1}{2} \left(\alpha  e^{-\epsilon }+\frac{\bar{\rho}_\mathrm{A}+\bar{\rho}_\mathrm{B}}{\tau }+\eta[\bar{\rho}_\mathrm{B}, \bar{\rho}_\mathrm{A}]  e^{-\epsilon }\right) 
    - \alpha \tau e^{ \epsilon }\frac{(\alpha\tau +(\bar{\rho}_\mathrm{A} + \bar{\rho}_\mathrm{B} )  e^{ \epsilon } ) \langle\!\langle \Delta \rho_\mathrm{A} (k, \omega) \Delta \rho_\mathrm{B} (-k, -\omega)\rangle\!\rangle}{(\alpha ^2 \tau ^2+2 \alpha  \tau  e^{\epsilon } (\bar{\rho}_\mathrm{A}+\bar{\rho}_\mathrm{B})+e^{2 \epsilon } (\bar{\rho}_\mathrm{A}-\bar{\rho}_\mathrm{B})^2)^{2}} \nonumber\\
    +\alpha \tau \, \eta[\bar{\rho}_\mathrm{B}, \bar{\rho}_\mathrm{A}]\, e^{2\epsilon }\frac{\left(\bar{\rho}_\mathrm{B} \langle\!\langle\Delta \rho_\mathrm{A} (k, \omega) \Delta \rho_\mathrm{A} (-k, -\omega) \rangle\!\rangle +  \bar{\rho}_\mathrm{A}\langle\!\langle\Delta \rho_\mathrm{B} (k, \omega) \Delta \rho_\mathrm{B} (-k, -\omega) \rangle\!\rangle  \right)}{(\alpha ^2 \tau ^2+2 \alpha  \tau  e^{\epsilon } (\bar{\rho}_\mathrm{A}+\bar{\rho}_\mathrm{B})+e^{2 \epsilon } (\bar{\rho}_\mathrm{A}-\bar{\rho}_\mathrm{B})^2)^{2}}
     \nonumber\\
    \label{eq:NumNetworked}
\end{eqnarray}  
\end{widetext}
Fig.~\ref{fig:numNetworkedPlotswithV}, shows the average number density of cross-links as given by ~eq.~\eqref{eq:NumNetworked}. The value is positive, as expected, and shows a similar shape to that of the individual structure factors. It should be noted here, that arbitrary units are used and that this quantity has not been normalised.
The cross-linked polymer mixture maintains stability, as depicted in ~Fig.~\ref{fig:NetworkedStructureFactorPlotswithV}, when including a sufficiently large repulsive term along with the dynamic effective potentials between polymers of the same type in the mixture. Investigating the analytical expressions for the dynamic structure factors (see Appendix \ref{app:corrFuns}), it can be shown that the dynamic structure factors do not diverge for $v \geq v_\mathrm{min}$ where
\begin{equation}
    v_\mathrm{min} = \tfrac{1}{2}\sqrt{w_\mathrm{A}^2 + 4 v_\mathrm{AB}^2 - 2 w_\mathrm{A} w_\mathrm{B} + w_\mathrm{B}^2}-\tfrac{1}{2}(w_\mathrm{A} +w_\mathrm{B})\,.
    \label{eq:Vmin}
\end{equation}

\begin{figure*}[htbp]
    \centering
    \subfigure[]{
        \includegraphics[width=0.45\textwidth]{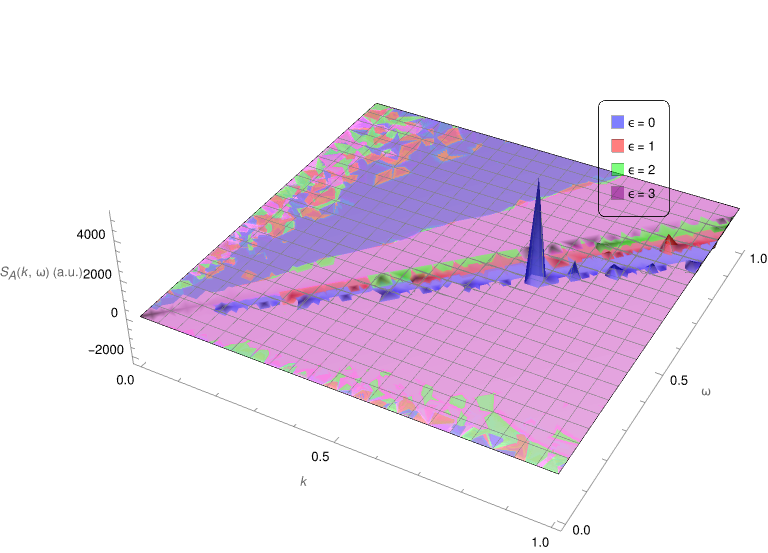}
        \label{fig:SAplot}
    }
    \subfigure[]{
        \includegraphics[width=0.45\textwidth]{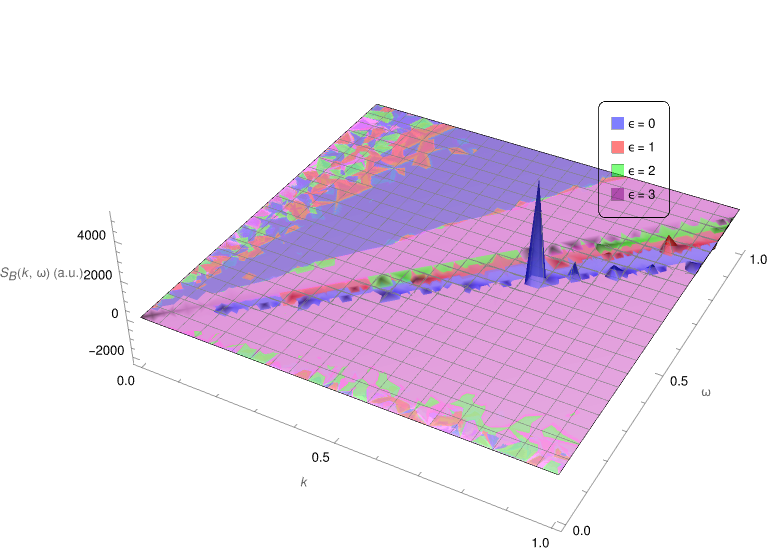}
        \label{fig:SBplot}
    }
    \subfigure[]{
        \includegraphics[width=0.45\textwidth]{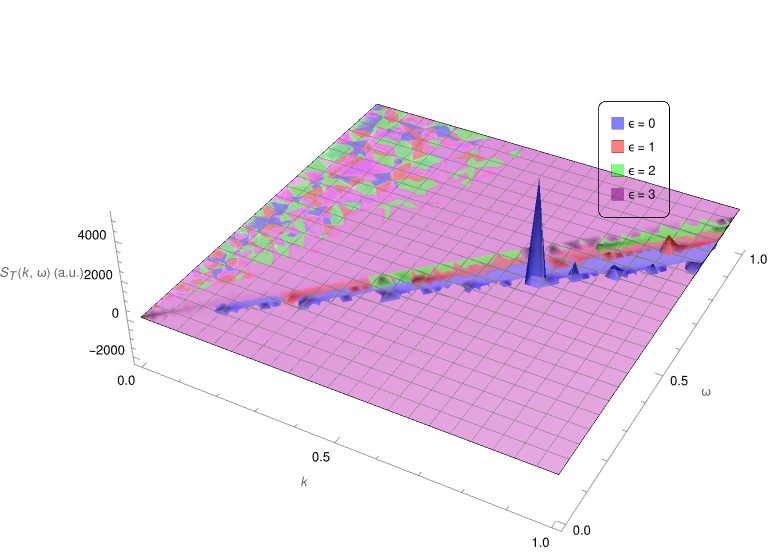}
        \label{fig:STplot}
    }
    \subfigure[]{
        \includegraphics[width=0.45\textwidth]{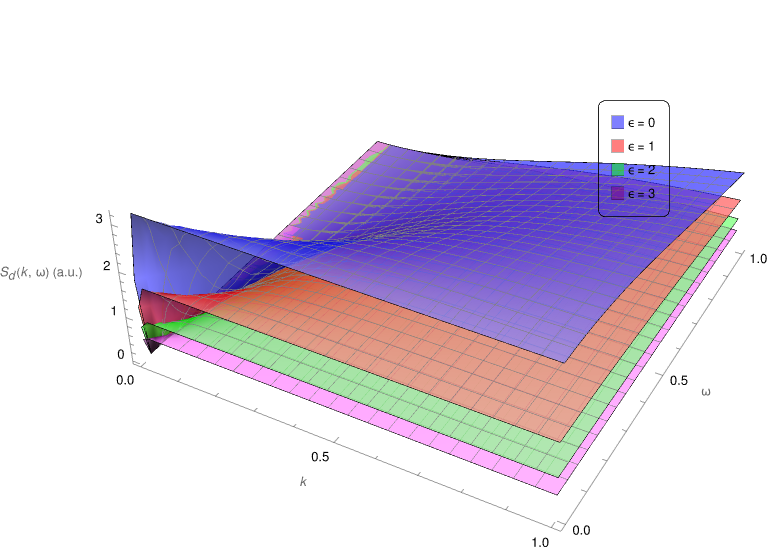}
        \label{fig:Sdplot}
    }
    \caption{Dynamic structure factors in arbitrary units for the cross-linked polymer solutions with  $\bar{\rho}_\mathrm{A} = \bar{\rho}_\mathrm{B} = 1$,$\gamma_\mathrm{A} = \gamma_\mathrm{B} =1$, $L_\mathrm{A}=L_\mathrm{B}=100$ and varying values of $\epsilon$. The upper panels, \ref{fig:SAplot} and \ref{fig:SBplot}, show the dynamic structure factors for the polymers with density $\bar{\rho}_\mathrm{A} + \Delta \rho_\mathrm{A} (k, \omega)$ and $\bar{\rho}_\mathrm{B} + \Delta \rho_\mathrm{B} (k, \omega)$, respectively. The lower panels, \ref{fig:STplot} and \ref{fig:Sdplot}, show the dynamic structure factors corresponding to the total polymer density and difference in polymer density as defined in ~eqs~\eqref{eq:ST}--\eqref{eq:Sd}.}
    \label{fig:NetworkedStructureFactorPlots}
\end{figure*}

Now, switching off the repulsive potential by setting $v=0$, the structure factors corresponding to this polymer mixture are shown in the panels of Fig.~\ref{fig:NetworkedStructureFactorPlots} for varying values of the networking or cross-linking advantage $\epsilon$. The upper panels Figs.~\ref{fig:SAplot}--\ref{fig:SBplot} show the structure factors corresponding to $\Delta \rho_\mathrm{A} (k, \omega)$ and $\Delta \rho_\mathrm{B} (k, \omega)$, respectively. The lower panels Figs.~\ref{fig:STplot}--\ref{fig:Sdplot}  show the structure factors corresponding to the fluctuations in the total polymer density $\Delta \rho_{T}(k, \omega)$ and the difference in densities $\Delta \rho_{d}(k, \omega)$, defined as follows:
\begin{subequations}
   \begin{eqnarray}
   \bar{\rho}_{T}+ \Delta \rho_{T}(k, \omega)= \tfrac{\bar{\rho}_\mathrm{A}+ \bar{\rho}_\mathrm{B}}{2} + \tfrac{\Delta{\rho}_\mathrm{A}(k, \omega)+ \Delta \rho_\mathrm{B}(k, \omega)}{2}\, \nonumber\\\label{eq:ST}\\
    \bar{\rho}_{d}+\Delta \rho_{d}(k, \omega)= \bar{\rho}_\mathrm{B}- \bar{\rho}_\mathrm{A}+\Delta \rho_\mathrm{B}(k, \omega)- \Delta \rho_\mathrm{A}(k, \omega) \, .\nonumber\\\label{eq:Sd}
    \end{eqnarray} 
    \label{eq:copolymerSdefs}
\end{subequations}

Of the four structure factors in the panels of Fig.~\ref{fig:NetworkedStructureFactorPlots}, it is only Fig.~\ref{fig:Sdplot}, the structure factor corresponding to $\Delta \rho_{d}(k, \omega)$, which displays the expected behaviour of a peak at low $k$ and $\omega$ values which drops off for higher  $k$ and $\omega$ values. The peak appears,as before, to grow lower and flatter as the networking advantage $\epsilon$ is increased. The behaviour seen in the remaining structure factors in Figs.~\ref{fig:SAplot}--\ref{fig:STplot} indicate a divergence of $S_\mathrm{A}(k, \omega)$ and $S_\mathrm{B}(k, \omega)$  due to their denominators equating to zero where, 
\begin{widetext}
 \begin{equation}
    \omega = \pm\sqrt{\tfrac{-\gamma_\mathrm{A} \gamma_\mathrm{B}2  w_\mathrm{A}-\gamma_\mathrm{A}2 \gamma_\mathrm{B} 2 w_\mathrm{B}\pm\sqrt{\gamma_\mathrm{A}^2 \gamma_\mathrm{B}^2  \left(\gamma_\mathrm{B}2 w_\mathrm{A}^2-2 \gamma_\mathrm{A} \gamma_\mathrm{B} w_\mathrm{A} w_\mathrm{B}+4 \gamma_\mathrm{A} \gamma_\mathrm{B} v_{AB}^2+\gamma_\mathrm{A}2 w_\mathrm{B}^2\right)}}{\gamma_\mathrm{A}^2 \gamma_\mathrm{B}^2}}\, k +  \mathcal{O}(k^3)\,.
    \label{eq:divergenceLine}
\end{equation}   
\end{widetext}
for small $k$. With this divergent behaviour also present where $k=0$ and $\omega=0$, \textit{i.e.}~in the long time limit and at large length scales,  this suggests a collapse of the polymer mixture.  Although this result is expected for a system that does not account for excluded volume interactions, since  attractive potentials are introduced, it remains useful to analyse this within the current approximation schemes. The structure factors originate from the Random Phase Approximation (RPA), therefore this divergence is an indication that the assumptions made during this approximation do not hold around the line given by eq.~\eqref{eq:divergenceLine}. The RPA assumes that the densities of the polymers are distributed amongst a homogeneous background density such that the fluctuations $\Delta \rho_\mathrm{A} (k, \omega)$ and $\Delta \rho_\mathrm{B} (k, \omega)$ around the background density remain sufficiently small. The divergence of the structure factors in Fig.~\ref{fig:SAplot}-\ref{fig:STplot} therefore indicate that the polymer mixture exhibits large fluctuations around the homogeneous background density without the inclusion of a sufficiently large repulsive potential.

\section{Conclusion}\label{sec:conclusion}
This paper introduces a novel field theoretical method for modelling dynamic networking in complex systems on the mesoscopic scale, focusing on systems where particles bind an unbind from one another. This modelling approach is built upon Langevin dynamics with the networking formalism acting as statistical weights selecting trajectories of particles that meet the networking conditions. The results  which are derivable within the formalism provide experimentally measurable quantities in a collective description.

The networking formalism can be set up in a variety of ways in order to represent different mechanisms of attachment or networking between particles and is therefore suitable for a broad range of applications. Constraints that specify the networking requirements are incorporated into pre-existing MSR generating functionals, such that this formalism can model dynamical networking in any system for which a dynamical structure factor can be obtained.

An example where networking is introduced into the dynamics of Brownian particles is used to illustrate the proposed approach and approximation schemes for implementing the networking formalism dynamically. This example highlights that the formalism can be interpreted on a discrete, microscopic level where the networking acts as weights for the selection of particle trajectories, as well as in a collective description, where the formalism acts via effective potentials between densities of networked particles. Since these potentials due to networking are attractive, this approach necessitates an additional repulsive potential. To conclude the illustrative example,the final generating functional with the networking and repulsive potentials is showcased along with a correlation function of the networked Brownian particles which demonstrate how the networking formalism can be used to couple the dynamics of two systems to one another.  From this correlation function a critical length scale is extracted, below which the dynamics appear purely diffusive, whilst above this length scale the dominant modes have slower relaxation times, indicating that the effects of networking dominates over the diffusive dynamics at these length scales.

An alternative implementation of the networking functional is derived with the addition of a networking advantage, which can be used as a scalable parameter for the strength or likelihood of cross-linking. This networking functional is utilised here to introduce cross-linking in a polymer mixture. This results in the polymer dynamics shifting from purely diffusive behaviour, without networking, towards less prominent diffusive peaks in the dynamic structure factors for increasing values of the networking advantage. The linear response functions in the long time limit show non-trivial behaviour on intermediate length scales which depends  on not only the polymer solution parameters but also the effective networking potentials. While there is no response to uniform fields in this limit, the response to local fields is consistent with that of the non-networked polymer solutions.

The intermittently cross-linked polymer mixture example also confirms that the current formalism leads to a collapse of the system if networking is introduced without the addition of a sufficiently large repulsive potential, as expected.  The collapse is indicated by a divergence of the dynamic structure factors for the cross-linked polymer mixture in the absence of a repulsive potential. Within the proposed formalism and approximation schemes, it is concluded that the assumption of small density fluctuations around a homogenous background density (introduced during the RPA), does not hold, therefore necessitating the incorporation of the repulsive potential to maintain stability of the networked system. 

The networking formalism and  proposed approximation schemes for implementing networking into a dynamical system are limited to dynamical systems where density fluctuations around a homogenous background are small. This limitation arises due to the use of the Random Phase Approximation to obtain dynamical structure factors for collective densities and is not inherent to the proposed networking formalism. The networking formalism itself necessitates an approximation scheme for the path integrals over the Gaussian fields, for which a saddle point approximation is proposed. The further small density fluctuation expansions are implemented merely to remain compatible with the dynamical systems in which networking is implemented. Should an alternative approach towards the modelling of the dynamical systems be identified, this current limitation could be circumvented in future implementations.

Future work could extend the cross-linking polymer mixture model by incorporating cross-linker particles which are themselves able to diffuse within the mixture, as opposed to cross-linking occurring directly between the beads of the polymers. This will introduce additional time and length-scales into the model which could lead to significant insights pertaining to how introducing cross-linking constraints on a microscopic level could cause interesting behaviour to emerge in the collective dynamics. The networking model could also be extended to include velocities in the MSR generating functional which are coupled to the density fields.  This would allow one to further investigate the effects of networking on the dynamics and linear response of the system. Alternatively, networking constraints can be imposed on velocities as well as positions of networked particles to implement longer-duration cross-links. These ideas could be particularly useful in systems which are expected to be viscoelastic, since when present, viscoelasticity can be identified in the velocity-velocity correlation functions. 

Although the limitations of the current approximation schemes make it difficult to model biological systems directly, the formalism and approximation schemes, as presented in this paper, could lead to physics based insights  pertaining to the dynamics of polymer networks which exhibit small density fluctuations. Further development of applicable approximation schemes is required to utilise the formalism to directly model non-homogenously distributed networks such as the cytoskeleton.

This novel approach for modelling dynamical networking has been rigorously studied through the illustrative examples presented in this paper, thereby paving the path for its future development and application to a variety of complex systems of biological and synthetic nature.

%\section*{Supplementary information}
%Not applicable.

\section*{Acknowledgements}

This work is based on the research supported
	
	in part 
	by the National Research Foundation of South Africa (Grant Numbers 99116, MND210620613719 and PMDS240820261081). \\
	%NdT acknowledges the financial support received from the National Research Foundation of South Africa through a postgraduate scholarship.\\
	KKMN would like to thank the Isaac Newton Institute for Mathematical Sciences, Cambridge, for support and hospitality during the programme ``New statistical physics in living matter: non equilibrium states under adaptive control'' where some of the work on this paper was undertaken. This work was supported by EPSRC grant no EP/R014604/1.

\section*{Author contribution}

%
%\begin{itemize}
%\item Funding: as per acknowledgements.
%\item Competing interests: The authors have no financial or proprietary interests in any material discussed in this article. 
%\item Ethics approval and consent to participate: Not applicable
%\item Consent for publication: Not applicable
%\item Data availability: Not applicable
%\item Materials availability: Not applicable
%\item Code availability: Not applicable
%\item Author contribution: 
Conceptualisation (NdT and KKMN), Calculations (detailed calculations and numerical work: NdT; approximation approach, checks: NdT and KKMN), Interpretation (NdT and KKMN), Writing (NdT), Editing (NdT and KKMN)
%\end{itemize}

\appendix
\section{Random Phase Approximation (RPA)}
\label{app:RPA_BM}
The functional integrals over the stochastic forces in eqs.~\eqref{eq:ZBM} may be evaluated such that the generating functionals are of the form:
\begin{equation}
Z =  \mathcal{N} \int [\mathrm{d} \mathbf{r}(t)]  [\mathrm{d} \hat{\mathbf{r}}(t)] \mathrm{e}^{\mathcal{F}[\mathbf{r}, \hat{\mathbf{r}}]}.
\label{eq:ZrpaStart}
\end{equation}

For ease of notation, here $J$ Brownian particles are considered which could represent either the set of beads or attachment points. --- each of which has a generating functional of the form of eq.~\eqref{eq:ZrpaStart}. The product of these, produces a generating functional for all $J$ Brownian particles:
\begin{multline}
Z =  \mathcal{N}\int [\mathrm{d} \mathbf{r}_1(t)] [\mathrm{d} \mathbf{r}_2(t)]...[\mathrm{d} \mathbf{r}_N(t)] [\mathrm{d} \hat{\mathbf{r}}_1(t)][\mathrm{d} \hat{\mathbf{r}}_2(t)]\\
...[\mathrm{d} \hat{\mathbf{r}}_N(t)] \mathrm{e}^{\sum_{j=1}^{J} \mathcal{F}[\mathbf{r}_j, \hat{\mathbf{r}}_j]}.
\label{eq:ZrpaStartN}
\end{multline}
For these $J$ Brownian particles, a collective density $\rho$ can be defined. Following \cite{fredricksonCollectiveDynamicsPolymer1990}, however, it is mathematically more convenient to utilise the spatial Fourier transform, $\rho(\mathbf{k}, \omega)$, of the concentration along with its corresponding auxiliary variable $\hat{\rho}(\mathbf{k}, \omega)$.

These variables are incorporated by multiplying the following into eq.~\eqref{eq:ZrpaStartN}
\begin{eqnarray}
    \int [\mathrm{d} \rho] [\mathrm{d} \hat{\rho}] \big\{ \delta(\rho - \sum_{j=1}^{J}\mathrm{e}^{i \mathbf{k} \cdot \mathbf{r}_j(t)})\nonumber\\
    \times \delta (\hat{\rho} - i\sum_{j=1}^{J}\mathbf{k} \cdot \mathbf{r}_j(t)\mathrm{e}^{i\mathbf{k} \cdot \mathbf{r}_j(t)}) \big\}\, .
    \label{eq:FH-collective-variables}
\end{eqnarray}
    This is equivalent to
    \begin{multline}
\mathcal{N} \int [\mathrm{d} \rho] [\mathrm{d} \hat{\rho}] [\mathrm{d} \psi_k] [\mathrm{d} \hat{\psi}_k]\mathrm{e}^{i \int_{k,t}\psi_k(\rho - \sum_{j=1}^{J}\mathrm{e}^{\mathrm{i} \mathbf{k} \cdot \mathbf{r}_j(t)})}\\
    \times \mathrm{e}^{\mathrm{i} \int_{k,t}\hat{\psi}_k(\hat{\rho} - \mathrm{i}\sum_{j=1}^{J} \mathbf{k} \cdot \mathbf{r}_j(t)\mathrm{e}^{\mathrm{i}  \mathbf{k} \cdot \mathbf{r}_j(t)})}.
    \end{multline}
Further utilising some second order expansions, the generating functional becomes:
\begin{widetext}
\begin{multline}
    Z = \mathcal{N}\int [\mathrm{d} \rho] [\mathrm{d} \hat{\rho}] [\mathrm{d} \psi_k] [\mathrm{d} \hat{\psi}_k]\mathrm{e}^{\mathrm{i} \int_{k,t}\psi_k \rho +i \int_{k,t}\hat{\psi}_k \hat{\rho}} \\
    \times \Bigg\{  \int [\mathrm{d} \mathbf{r}_1(t)] [\mathrm{d} \mathbf{r}_2(t)]...[\mathrm{d} \mathbf{r}_N(t)] [\mathrm{d} \hat{\mathbf{r}}_1(t)][\mathrm{d} \hat{\mathbf{r}}_2(t)]...[\mathrm{d} \hat{\mathbf{r}}_N(t)]\\
    \times  \mathrm{e}^{\sum_{j=1}^{J} \mathcal{F}[\mathbf{r}_j, \hat{\mathbf{r}}_j]} \left(1-  \mathrm{i} \int_{k,t}\psi_k \sum_{j=1}^{J} \mathrm{e}^{\mathrm{i} \mathbf{k} \cdot \mathbf{r}_j(t)}\right.\\ \left.+ \int_{k,t}\hat{\psi}_k \sum_{j=1}^{J} \mathbf{k} \cdot \hat{\mathbf{r}}_j(t)\mathrm{e}^{\mathrm{i} \mathbf{k} \cdot\mathbf{r}_j(t)} \right. \\ \left.- \mathrm{i} \int_{k,t}\int_{k',t'} \psi_k \hat{\psi}_{k'} \sum_{j=1}^{J} \sum_{\alpha=1}^{J} \mathbf{k'} \cdot \hat{\mathbf{r}}_\alpha (t)\mathrm{e}^{\mathrm{i} \mathbf{k} \cdot \mathbf{r}_j(t) + \mathrm{i} \mathbf{k'} \cdot \mathbf{r}_\alpha(t')} \right. \\ \left.+\frac{1}{2} \int_{k,t}\int_{k',t'} \hat{\psi}_k \hat{\psi}_{k'} \sum_{j=1}^{J} \sum_{\alpha=1}^{J} \mathbf{k} \cdot \hat{\mathbf{r}}_j (t)\, \mathbf{k'} \cdot \hat{\mathbf{r}}_\alpha (t')\mathrm{e}^{\mathrm{i} \mathbf{k} \cdot \mathbf{r}_j(t) + \mathrm{i} \mathbf{k'} \cdot \mathbf{r}_\alpha(t')} 
    \right. \\ 
    \left.+\frac{1}{2} \int_{k,t}\int_{k',t'} \psi_k \psi_{k'} \sum_{j=1}^{J} \sum_{\alpha=1}^{J} \mathrm{e}^{\mathrm{i} \mathbf{k} \cdot \mathbf{r}_j(t) + \mathrm{i} \mathbf{k'} \cdot \mathbf{r}_\alpha(t')}\right)\Bigg\}.
\end{multline}
\end{widetext}
At this point, the functional integrals over the $\mathbf{r}_j(t)$ and $\hat{\mathbf{r}}_j(t)$ are evaluated for each of the terms in the expansion. 

One of these functional integrals merely results in the average concentration of motor proteins. The crux of this approximation is that this average concentration, or $k=0$, term may be omitted.

After some final mathematical manipulation of the remaining functional integrals, one obtains:

\begin{multline}
    Z = \mathcal{N}\int [\mathrm{d} \rho] [\mathrm{d} \hat{\rho}] [\mathrm{d} \psi_k] [\mathrm{d} \hat{\psi}_k]\Bigg\{\mathrm{e}^{\mathrm{i} \int_k \int_\omega\psi_k \rho} \\
        \times \mathrm{e}^{\mathrm{i} \int_k \int_\omega \hat{\psi}_k \hat{\rho}} \left(1- \frac{1}{2} \int_k \int_\omega  \psi_k(\omega) S(\mathbf{k}, \omega) \psi_{-k}(-\omega)\right.\\ \left. +\mathrm{i} \int_k \int_\omega  \psi_k(\omega) \chi (-\mathbf{k},-\omega) \hat{\psi}_{-k}(-\omega) \right)\Bigg\}
\end{multline}

Moving everything back into the exponent yields
\begin{multline}
    Z_\mathrm{RPA} = \mathcal{N} \int [\mathrm{d} \rho] [\mathrm{d}\hat{\rho}]   \mathrm{e}^{\mathrm{i} \int_k \int_\omega  \rho (\mathbf{k},\omega) \chi^{-1} (-\mathbf{k},-\omega) \hat{\rho} (-\mathbf{k},-\omega) }\\
    \times\,\mathrm{e}^{-\frac{1}{2} \int_k \int_\omega \hat{\rho} (\mathbf{k},\omega) \chi^{-1}(\mathbf{k},\omega) S(\mathbf{k}, \omega)\chi^{-1} (-\mathbf{k},-\omega)  \hat{\rho} (-\mathbf{k},-\omega)}, 
    \label{eq:ZRPA}
\end{multline}

implementing the Gaussian functional integral over $\hat{\rho}$, one arrives at eqs.~\eqref{eq:ZBMcollective}.

\section{Correlation and linear response functions}
\label{app:corrFuns}
One can combine the generating functionals for two dynamical systems with densities $\rho_\mathrm{A}$ and $\rho_\mathrm{B}$ using a networking functional, as done in eq.~\eqref{eq:Zfull} , with the generating functional describing the dynamics of the system (see eq.~\eqref{eq:ZRPA}) to obtain
\begin{widetext}
\begin{eqnarray}
   \mathbb{Z}= \mathcal{N} \int [\mathrm{d} \Delta \rho_\mathrm{A}]  [\mathrm{d} \Delta \rho_\mathrm{B}] [\mathrm{d} \Delta \hat{\rho}_\mathrm{A}]  [\mathrm{d} \Delta \hat{\rho}_\mathrm{B}]   \mathrm{e}^{\mathrm{i} \int_k \int_\omega   \Delta\rho_\mathrm{A} (\mathbf{k},\omega) \chi_{0,\mathrm{A}}^{-1} (-\mathbf{k},-\omega) \Delta \hat{\rho}_\mathrm{A} (-\mathbf{k},-\omega)  +\mathrm{i} \int_k \int_\omega   \Delta\rho_\mathrm{B} (\mathbf{k},\omega) \chi_{0,\mathrm{B}}^{-1} (-\mathbf{k},-\omega) \Delta \hat{\rho}_\mathrm{B} (-\mathbf{k},-\omega) }\nonumber\\
    \times\,\mathrm{e}^{-\frac{1}{2} \int_k \int_\omega \Delta \hat{\rho}_\mathrm{A} (\mathbf{k},\omega) \chi_{0,\mathrm{A}}^{-1}(\mathbf{k},\omega) S_{0,\mathrm{A}}(\mathbf{k}, \omega)\chi_{0,\mathrm{A}}^{-1} (-\mathbf{k},-\omega)  \Delta \hat{\rho}_\mathrm{A} (-\mathbf{k},-\omega)-\frac{1}{2} \int_k \int_\omega \Delta \hat{\rho}_\mathrm{B} (\mathbf{k},\omega) \chi_{0,\mathrm{B}}^{-1}(\mathbf{k},\omega) S_{0,\mathrm{B}}(\mathbf{k}, \omega)\chi_{0,\mathrm{B}}^{-1} (-\mathbf{k},-\omega)  \Delta \hat{\rho}_\mathrm{B} (-\mathbf{k},-\omega)}\nonumber\\
    \times \mathrm{e}^{-\int_{\mathbf{k}, \omega} \Delta \rho_\mathrm{A}(\mathbf{k}, \omega) \,v_{AB}\, \Delta \rho_\mathrm{B}(-\mathbf{k}, -\omega)-\frac{1}{2 } \int_{\mathbf{k}, \omega} \Delta \rho_\mathrm{A}(\mathbf{k}, \omega) \,( w_\mathrm{A}+v)\, \Delta \rho_\mathrm{A}(-\mathbf{k}, -\omega)-\frac{1}{2 } \int_{\mathbf{k}, \omega} \Delta \rho_\mathrm{B}(\mathbf{k}, \omega) \,(w_\mathrm{B}+v)\, \Delta \rho_\mathrm{B}(-\mathbf{k}, -\omega)}\, \, \nonumber\\
    \times   \mathrm{e}^{\mathrm{i}\int_{\mathbf{k}, \omega}  J_\mathrm{A}(\mathbf{k}, \omega ) \,\Delta \rho_\mathrm{A}(-\mathbf{k}, -\omega)+\mathrm{i}\int_{\mathbf{k}, \omega}  J_\mathrm{B}(\mathbf{k}, \omega ) \,\Delta \rho_\mathrm{B}(-\mathbf{k}, -\omega)+\mathrm{i}\int_{\mathbf{k}, \omega}  \hat{J}_\mathrm{A}(\mathbf{k}, \omega ) \,\Delta \hat{\rho}_\mathrm{A}(-\mathbf{k}, -\omega)+\mathrm{i}\int_{\mathbf{k}, \omega}  \hat{J}_\mathrm{B}(\mathbf{k}, \omega ) \,\Delta \hat{\rho}_\mathrm{B}(-\mathbf{k}, -\omega)}. \nonumber \\
\end{eqnarray}
\end{widetext}

Here $S_{0,\mathrm{A}}(\mathbf{k}, \omega)$ and $S_{0,\mathrm{B}}(\mathbf{k}, \omega)$ are the dynamical structure factors of the non-networked systems and $\chi_{0,\mathrm{A}}(\mathbf{k}, \omega)$ and $\chi_{0,\mathrm{B}}(\mathbf{k}, \omega)$ are their linear response functions, whilst $w_\mathrm{A}, w_\mathrm{B} \text{and} v_{AB}$ are the magnitudes of the effective potentials due to networking, in this paper given either by eqs.~\eqref{eq:effW_B}--\eqref{eq:effV_AB} or eqs.~\eqref{eq:wBAdv}--\eqref{eq:v_ABAdv}.  An additional repulsive potential $v$ is also added between particles of the same type is included to maintain stability of the system with $v\geq\tfrac{1}{2}\sqrt{w_\mathrm{A}^2 + 4 v_\mathrm{AB}^2 - 2 w_\mathrm{A} w_\mathrm{B} + w_\mathrm{B}^2}-\tfrac{1}{2}(w_\mathrm{A} +w_\mathrm{B})\,$. 

After evaluating the Gaussian functional integrals, one can take the partial derivatives with respect to $J_\mathrm{A}$ and $J_\mathrm{B}$ as in eq.~\eqref{eq:aveFormulas} to obtain the following expressions for the correlation and cross-correlation functions of $\Delta \rho_\mathrm{A}$ and $\Delta \rho_\mathrm{B}$

\begin{widetext}
\begin{equation}
   \langle\!\langle\,  \Delta \rho_\mathrm{A} ( \mathbf{k}, \omega) \Delta \rho_\mathrm{A} ( -\mathbf{k},- \omega)\,\rangle\!\rangle =\frac{S_{0,\mathrm{A}}( \mathbf{k}, \omega)\left(1+ S_{0,\mathrm{B}}( \mathbf{k}, \omega)\,( w_\mathrm{B}+v)\right)}{1 + S_{0,\mathrm{B}}( \mathbf{k}, \omega) \,( w_\mathrm{B}+v) + S_{0,\mathrm{A}}( \mathbf{k}, \omega) \left(w_\mathrm{A} + v+S_{0,\mathrm{B}}( \mathbf{k}, \omega) \left((w_\mathrm{A}+v)( w_\mathrm{B}+v)-  v_{AB}^2\right) \right)}\, ,
    \label{eq:AcorrNew}
\end{equation}
\begin{equation}
   \langle\!\langle\,  \Delta \rho_\mathrm{B} ( \mathbf{k}, \omega) \Delta \rho_\mathrm{B} ( -\mathbf{k},- \omega)\,\rangle\!\rangle =\frac{S_{0,\mathrm{B}}( \mathbf{k}, \omega)\left(1 + S_{0,\mathrm{A}}( \mathbf{k}, \omega) \,(w_\mathrm{A}+v)\right)}{1 + S_{0,\mathrm{B}}( \mathbf{k}, \omega)\, (w_\mathrm{B} +v)+ S_{0,\mathrm{A}}( \mathbf{k}, \omega) \left(w_\mathrm{A}+v + S_{0,\mathrm{B}}( \mathbf{k}, \omega) \left((w_\mathrm{A}+v)( w_\mathrm{B}+v)-  v_{AB}^2\right) \right)}\, ,
    \label{eq:BcorrNew}
\end{equation}
\begin{equation}
   \langle\!\langle\,  \Delta \rho_\mathrm{A} ( \mathbf{k}, \omega) \Delta \rho_\mathrm{B} ( -\mathbf{k},- \omega)\,\rangle\!\rangle =\frac{S_{0,\mathrm{A}}( \mathbf{k}, \omega) S_{0,\mathrm{B}}( \mathbf{k}, \omega) \, v_{AB}}{1 + S_{0,\mathrm{B}}( \mathbf{k}, \omega)\, (w_\mathrm{B} +v)+ S_{0,\mathrm{A}}( \mathbf{k}, \omega) \left(w_\mathrm{A}+v + S_{0,\mathrm{B}}( \mathbf{k}, \omega) \left((w_\mathrm{A}+v)( w_\mathrm{B}+v)-  v_{AB}^2\right) \right)}\,.
    \label{eq:ABcorrNew}
\end{equation}
\end{widetext}
In addition to $J_\mathrm{A}$ and $J_\mathrm{B}$, also taking partial derivatives with respect to $\hat{J}_\mathrm{A}$ and $\hat{J}_\mathrm{B}$ , allows one to obtain the follow expressions for the linear response functions
\begin{widetext}
\begin{equation}
   \langle\!\langle\,  \Delta \rho_\mathrm{A} ( \mathbf{k}, \omega) \Delta \hat{\rho}_\mathrm{A} ( -\mathbf{k},- \omega)\,\rangle\!\rangle =\frac{\chi_{0,\mathrm{A}}( \mathbf{k}, \omega)(S_{0,\mathrm{B}}( \mathbf{k}, \omega) (v+w_\mathrm{B})+1)}{1 + S_{0,\mathrm{B}}( \mathbf{k}, \omega) \,( w_\mathrm{B}+v) + S_{0,\mathrm{A}}( \mathbf{k}, \omega) \left(w_\mathrm{A} + v+S_{0,\mathrm{B}}( \mathbf{k}, \omega) \left((w_\mathrm{A}+v)( w_\mathrm{B}+v)-  v_{AB}^2\right) \right)}\, ,
    \label{eq:AresponseNew}
\end{equation}
\begin{equation}
   \langle\!\langle\,  \Delta \rho_\mathrm{B} ( \mathbf{k}, \omega) \Delta\hat{ \rho}_\mathrm{B} ( -\mathbf{k},- \omega)\,\rangle\!\rangle =\frac{\chi_{0,\mathrm{B}}( \mathbf{k}, \omega)(S_{0,\mathrm{A}}( \mathbf{k}, \omega) (v+w_\mathrm{A})+1)}{1 + S_{0,\mathrm{B}}( \mathbf{k}, \omega) \,( w_\mathrm{B}+v) + S_{0,\mathrm{A}}( \mathbf{k}, \omega) \left(w_\mathrm{A} + v+S_{0,\mathrm{B}}( \mathbf{k}, \omega) \left((w_\mathrm{A}+v)( w_\mathrm{B}+v)-  v_{AB}^2\right) \right)}\, ,
    \label{eq:BresponseNew}
\end{equation}
\begin{equation}
   \langle\!\langle\,  \Delta \rho_\mathrm{A} ( \mathbf{k}, \omega) \Delta\hat{ \rho}_\mathrm{B} ( -\mathbf{k},- \omega)\,\rangle\!\rangle =-\frac{S_{0,\mathrm{A}}( \mathbf{k}, \omega) v_\mathrm{AB} \text{$\chi $B}}{1 + S_{0,\mathrm{B}}( \mathbf{k}, \omega) \,( w_\mathrm{B}+v) + S_{0,\mathrm{A}}( \mathbf{k}, \omega) \left(w_\mathrm{A} + v+S_{0,\mathrm{B}}( \mathbf{k}, \omega) \left((w_\mathrm{A}+v)( w_\mathrm{B}+v)-  v_{AB}^2\right) \right)}\, ,
    \label{eq:ABresponseNew}
\end{equation}
\begin{equation}
   \langle\!\langle\,  \Delta \rho_\mathrm{B} ( \mathbf{k}, \omega) \Delta\hat{ \rho}_\mathrm{A} ( -\mathbf{k},- \omega)\,\rangle\!\rangle =-\frac{S_{0,\mathrm{B}}( \mathbf{k}, \omega) v_\mathrm{AB} \text{$\chi $A}}{1 + S_{0,\mathrm{B}}( \mathbf{k}, \omega) \,( w_\mathrm{B}+v) + S_{0,\mathrm{A}}( \mathbf{k}, \omega) \left(w_\mathrm{A} + v+S_{0,\mathrm{B}}( \mathbf{k}, \omega) \left((w_\mathrm{A}+v)( w_\mathrm{B}+v)-  v_{AB}^2\right) \right)}\, .
    \label{eq:BAresponseNew}
\end{equation}
\end{widetext}

\section{Fluctuations around the Saddle Point}
\label{app:SPforAverage}
Utilising the quantity $\alpha \bar{\Phi}(\mathbf{r},t) \bar{\Phi}^* (\mathbf{r},t)$ as a measure of the average number of instance of networking would in fact require the calculation of the following average, or integral over the fields $\Phi$ and $\Phi^*$
\begin{equation}
   \langle \alpha  \Phi(\mathbf{r},t) \Phi^* (\mathbf{r},t) \rangle=\alpha  \int [\mathrm{d} \Phi][\mathrm{d} \Phi^*] \Phi(\mathbf{r},t) \Phi^* (\mathbf{r},t) \mathrm{e}^{\mathbb{F}[\Phi,\Phi^*]} \, .
    \label{eq:networkedAve}
\end{equation}
Rewriting \eqref{eq:networkedAve} such that everything is in the exponent,
\begin{equation}
    \langle \alpha  \Phi(\mathbf{r},t) \Phi^* (\mathbf{r},t)\rangle =\alpha  \int [\mathrm{d} \Phi][\mathrm{d} \Phi^*]  \mathrm{e}^{\mathrm{ln}(\Phi(\mathbf{r},t) \Phi^* (\mathbf{r},t)) +\mathbb{F}[\Phi,\Phi^*]} \, ,
    \label{eq:networkingAveExp}
\end{equation}
it becomes clear that a new saddle point solution should be determined in order to approximate the functional integrals over $\Phi$ and $\Phi^*$. Let  $\bar{\Phi}_\mathrm{new}$ and $\bar{\Phi}^*_\mathrm{new}$ denote the saddle point solutions for ~eq.~\eqref{eq:networkingAveExp}. Omitting the dependence on $\mathbf{r}$ and $t$ for ease of writing, let 
\begin{eqnarray}
    \bar{\Phi}_\mathrm{new} =\bar{\Phi} + \Delta \bar{\Phi} \\
    \bar{\Phi}^*_\mathrm{new} =\bar{\Phi}^* + \Delta \bar{\Phi}^* 
\end{eqnarray}
then, the saddle point solution for ~eq.~\eqref{eq:networkingAveExp} can be written as
\begin{multline}
    \bar{\Phi}_\mathrm{new} \bar{\Phi}^*_\mathrm{new}\mathrm{e}^{\mathbb{F}[\bar{\Phi}_\mathrm{new}, \bar{\Phi}^*_\mathrm{new}]} = (\bar{\Phi} \bar{\Phi}^* + \bar{\Phi} \Delta \Phi^* \\+ \bar{\Phi}^* \Delta \Phi +\Delta \Phi \Delta \Phi^*)\mathrm{e}^{\mathbb{F}[\bar{\Phi} + \Delta \bar{\Phi} ,\bar{\Phi}^* + \Delta \bar{\Phi}^* ]}\, .
\end{multline}
Now, expanding up to first order in the fluctuations $\Delta \bar{\Phi}$ and $\Delta \bar{\Phi}^* $ around the original saddle point, we find
\begin{equation}
    \mathbb{F}[\bar{\Phi} + \Delta \bar{\Phi} ,\bar{\Phi}^* + \Delta \bar{\Phi}^* ] = \mathbb{F}[\bar{\Phi} ,\bar{\Phi}^* ] + \Delta \mathbb{F}
\end{equation}
where
\begin{multline}
    \Delta \mathbb{F} = \frac{1}{\tau} \int_{\mathbf{r},t} \frac{\rho_\mathrm{A}}{1 + \bar{\Phi}}\Delta \Phi+  \frac{1}{\tau} \int_{\mathbf{r},t} \frac{\rho_\mathrm{B} \mathrm{e}^\epsilon}{1+\mathrm{e}^\epsilon\bar{\Phi}^*}\Delta \Phi^* \\
    - \alpha \int_{\mathbf{r},t}(\bar{\Phi} \Delta \Phi^* + \bar {\Phi}^* \Delta \Phi)\, 
\end{multline}
Recalling  eqs.~\eqref{eq:SP1} and \eqref{eq:SP2}, it is evident that in this first order expansion, $\Delta \mathbb{F}=0$. Thus, neglecting terms that are second order in the fluctuations, the saddle point solution for ~eq.~\eqref{eq:networkingAveExp} can be written as
\begin{equation}
        \bar{\Phi}_\mathrm{new} \bar{\Phi}^*_\mathrm{new}\mathrm{e}^{\mathbb{F}[\bar{\Phi}_\mathrm{new}, \bar{\Phi}^*_\mathrm{new}]} = \bar{\Phi} \bar{\Phi}^*\mathrm{e}^{\mathbb{F}[\bar{\Phi}  ,\bar{\Phi}^* ]} + (\bar{\Phi} \Delta \Phi^* + \bar{\Phi}^* \Delta \Phi)\mathrm{e}^{\mathbb{F}[\bar{\Phi}  ,\bar{\Phi}^* ]}\, .
\end{equation}
From this result, it will be assumed that the original saddle point solutions $\bar{\Phi}$ and $\bar{\Phi}^*$ serve as a sufficient approximation for the purpose of calculating the average number of networked particles according to ~eq.~\eqref{eq:numberNetworkedAve}. Thus, the saddle point solutions can merely be substituted into this expression to obtain the relevant average in terms of the collective variables $\rho_\mathrm{B}(\mathbf{r},t)$ and $ \rho_\mathrm{A} ( \mathbf{r},t)$.

\bibliographystyle{apsrev4-2} 
%\bibliography{references}
\input{main.bbl}
\end{document}